\documentstyle[aps,preprint,epsfig]{revtex}

\tightenlines

\begin{document}

\title{Collapsing Bacterial Cylinders}

\author{M. D. Betterton$^1$ and Michael P. Brenner$^2$}
\address{$^1$ Department of Physics, Harvard University, Cambridge, MA 02138 \\
$^2$ Department of Mathematics, MIT, Cambridge, MA 02139}

\maketitle

\begin{abstract}
Under special conditions bacteria excrete an attractant and
aggregate. The high density regions initially collapse into
cylindrical structures, which subsequently destabilize and break up
into spherical aggregates.  This paper presents a theoretical
description of the process, from the structure of the collapsing
cylinder to the spacing of the final aggregates. We show that
cylindrical collapse involves a delicate balance in which bacterial
attraction and diffusion nearly cancel, leading to
corrections to the collapse laws expected from dimensional analysis.
The instability of a collapsing cylinder is composed of two distinct
 stages: Initially, slow modulations to the cylinder develop, which
correspond to a variation of the collapse time along the cylinder
axis.  Ultimately, one point on the cylinder
pinches off.  At this final stage of the instability, a front
propagates from the pinch into the remainder of the cylinder.  The
spacing of the resulting spherical aggregates is determined by the
front propagation.
\end{abstract}
\newpage

The formation of a singularity---the divergence of a physical quantity
in finite time---is central to diverse fields\cite{cafl},
including nonlinear optics, gravitational collapse, and fluid
mechanics. The structure of singularities has been worked out in many
examples for which a physical quantity blows up at a spatial
point\cite{pumsig,egg,lars}.
Typically, singular dynamics are self-similar: the characteristic scale
separation between the singular and regular parts of the solution
leads to the slaving of the spatial structure  to
the time dependence via scaling laws.  The situation can be more
complicated when many singularities form collectively and
simultaneously. In this paper we analyze a simple example for which
multiple singularities form in a short time.  This work was
motivated by a recent experiment in bacterial
chemotaxis\cite{bud91,bud95,bre98}.

\begin{figure}[p]
\centerline{\epsfysize=2.5in\epsfbox{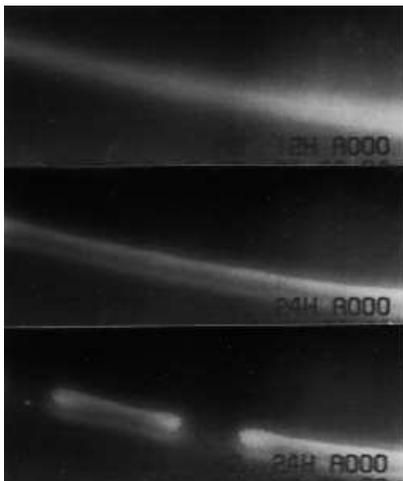}}
\caption[]
{Experiment showing formation and instability of
a collapsing bacterial cylinder (Reproduced from \cite{bre98}).
The first panel shows
a diffuse cloud of bacteria filling the depth of a petri dish
filled with agar, which then collapses (second panel) into
a cylindrical structure.  The cylinder subsequently destablizes
into spherical aggregates.  Details of the experiments are described
in Budrene and Berg \cite{bud91,bud95}. }
\label{elena}
\end{figure}

The experimental observation is shown in Figure \ref{elena}.  In the
first panel  a diffuse cloud of {\it Escherichia coli} covers
the depth of a petri dish filled with agar. The environment is
prepared so that the {\it E. coli} excrete an attractant; each
bacterium attracts all the other bacteria, and a cloud can collapse.
In the second panel, the diffuse cloud collapses as a cylindrical
structure, with highest bacterial density on the cylinder axis.  In
the final panel, the cylinder breaks down into spherical
aggregates. In this paper we analyze this process, by constructing a
similarity solution to describe the cylindrical collapse of bacteria,
and then analyzing its stability.

Chemotaxis in {\it E. coli} is an excellent model
 system for studying singularity formation.  The
biochemical response of {\it E. coli} to a changing 
environment has been thoroughly characterized
\cite{ber90,wol89,blo84,bla88,seg86} over the past twenty-five years,
so we understand how the bacteria sense and respond to their
environment.  As a consequence,
it is possible to write down a ``first principles'' hydrodynamic
theory for the motion of many bacteria
\cite{sch90,sch93} in which
 the response coefficients are measurable. Quantitative comparison
 between theory and experiment is possible, and any discrepancies can
 be traced directly to the biochemistry of individual
 bacteria\cite{bre98}.  The application to singularity formation arose
 from the recent discovery by Budrene and Berg \cite{bud91,bud95} of
 an assay in which {\it E. coli} excrete aspartate, an amino acid
 which is also an attractant for nearby bacteria.  Attractant
 diffusion drives aggregation because it leads to an effective force
 between individual bacteria: a higher density of bacteria in a given
 region leads to a higher attractant concentration, which then
 attracts more bacteria.

The initial interest in the Budrene-Berg experiment was stimulated by
the symmetrical patterns that form when chemotactic bacteria are
seeded in the center of a petri dish, as shown in their papers
\cite{bud91,bud95}. 

Several theories have been developed for these patterns, most of which
\cite{tys96,tys97,ben95,tsi95,woo95,bru92} view the pattern formation
as resulting from a linear instability of a (1 dimensional) travelling
wave of bacteria.  Recently, it was pointed out \cite{bre98} that each
of the aggregates in a a pattern corresponds to a density singularity
in the hydrodynamic description of the bacteria.  Therefore the
pattern formation depends crucially on the dynamics of singularity
formation.  Singularities in chemotaxis were anticipated by
Nanudjiah\cite{nan73} and Childress and Percus\cite{chi81} in studies
of mathematical models of chemotaxis.  An important feature,
understood first by Childress and Percus, is that chemotactic collapse
has a {\sl critical dimension}: although collapse to an infinite
density sheet is mathematically impossible, collapse to infinite
density lines and points both can occur.  It was argued in
\cite{bre98} that these facts crucially affect the patterns that can
form .

In particular, Figure 1 shows a step in the formation of the
aggregates.  The initially diffuse band (filling the depth of agar)
cannot form a singularity by collapsing only one of its dimensions to
zero thickness; instead it collapses into a cylinder (contracting two
of its dimensions simultaneously). The cylinder later destabilizes to
form aggregates, for which all three dimensions contract
simultaneously.  Models \cite{tys96,tys97,ben95,tsi95,woo95,bru92}
viewing aggregate formation as the linear instability of a band cannot
account for these experimental observations. These two different
pictures of aggregates form lead to different conclusions about which
biochemical parameters set the wavelength and structure of the
patterns. For the ``collapsing cylinder'' mechanism advocated here,
the characteristics of the pattern are set by the same biochemical
 cutoff which prevents an aggregate from reaching infinite density.

Cylindrical collapse is also important when a {\sl uniform} density
cloud of bacteria breaks into aggregates.  Linear stability analysis
of the uniform density state predicts that the cloud directly breaks
down into spherical aggregates.  However, experiments\cite{budper}
find that the the clumping is hierarchical: the uniform density cloud
first collapses as cylindrical structures which then break into
spherical aggregates. An important unsolved question is to explain the
geometry of the high density regions during collapse, and to predict
the distribution of final aggregates.

\begin{figure}[p]
\centerline{\epsfysize=5in\epsfbox{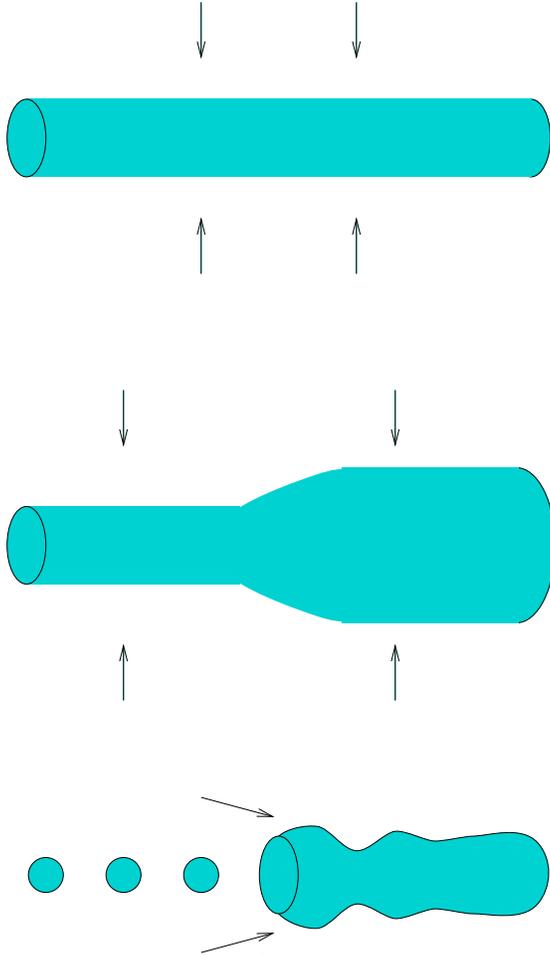}}
\caption[]
{Schematic of the three phases of cylindrical collapse. Top, a
collapsing cylinder is formed. Middle, the cylinder becomes
modulated. Bottom, the modulated cylinder pinches off and contracts,
leaving a series of spherical aggregates. Note this sketch does not
show the radial contraction of the cylinder which takes place as it
collapses.
\label{cylsketch}}
\end{figure}

In this paper, we use a combination of simulations and asymptotics to
describe the breakdown of a cylinder in of three principal
steps (Figure \ref{cylsketch}).  First, the bacteria collapse as a
cylinder towards a line of infinite density.  In the second step,
uniformity along the cylinder axis is broken, and a singularity
develops at a single point.  Finally, the remaining cylinder breaks up
producing a sequence of spherical aggregates.  Our primary conclusion
connecting the present theory with the experiments is that the spacing
between the aggregates in patterns such as Figure 2 is determined by
the local depletion of chemicals which make aspartate production
possible.  According to Budrene\cite{budper}, the most likely
candidate for this is the overhead oxygen concentration in the
cell. This prediction is qualitatively in accord with the experiments;
moreover this parameter dependence could be directly tested in future
experiments, and would serve to discriminate this theory from those
based on pure linear stability analysis.

In the following section, we review the basics of
chemotaxis, and discuss details necessary to understand the
Budrene-Berg experiments.  Section 2 describes the cylindrical
collapse of the bacteria.  Cylindrical collapse (in which two
dimensions of the cloud contract simultaneously) is a critical case
\cite{chi81,chi84}, for which diffusion and attraction nearly exactly
balance.  This criticality has complicated previous attempts to solve the collapse; a recent treatment of collapse in the 2D nonlinear Schrodinger 
equation \cite{pap99} inspired our solution.  In Section 3 we perform
a stability analysis of a collapsing cylinder. Perturbations to
the  cylinder can be described by a ``phase equation'' for the
singular time\cite{manneville}.  Solutions to this envelope
equation explain full numerical simulations of a modulated cylinder.
Section 4 describes the final stage of the breakdown of the cylinder,
after the cylinder has pinched off at a point. Stability analysis 
predicts the breakup of the remaining column of bacteria, a situation 
analogous to a propagating Rayleigh instability
in a liquid column\cite{pow98}. Appendices describe our numerical
methods, and fill in the details of the calculations.

\section{Problem Formulation: Bacterial Chemotaxis}

Chemotaxis refers to the migration of bacteria up chemical gradients.
For {\it Escherichia coli}, the basis of chemotaxis is largely
understood \cite{ber88,stock96}: in the absence of a chemical
gradient, an {\it E. coli} bacterium performs a random
walk\cite{ber72}.  When chemical gradients are present, the
bacterium's internal biochemical reactions detect the gradients and
couple to the bacterial movement system. This sensing biases the
random walk, and the bacterium has a net drift towards a chemical
attractant.  Under special conditions, the bacteria can excrete the
chemoattractant aspartate \cite{bud91,bud95} by converting carbon and
nitrogen sources (succinate and ammonia, respectively) in their
environment.  In these experiments no {\sl external} chemical
gradients are present.  Instead, each bacterium moves in response to
the attractant produced by other bacteria.  Thus, the excretion of
attractant produces a long-range force between the bacteria, and
induces complicated interactions in the colony.

The equations for the collective motion of the bacteria can be derived
(with no free parameters) from the underlying
biochemistry\cite{sch93}, allowing quantitative comparisons between
theory and experiments.  The basic equations for the bacterial density
$\rho$ and the attractant concentration $c$ are
\begin{eqnarray}
\frac{\partial \rho}{\partial t} &=& D_b \nabla^2 \rho 
-  \nabla\cdot\biggl(k \rho \nabla
c\biggr) +  a \rho \label{ks1}\\
\frac{\partial c}{\partial t} &= &D_c \nabla^2 c + \alpha \rho \label{ks2}.
\end{eqnarray}
Here $D_b$ is the bacterial diffusion constant, $k$ the chemotactic
coefficient, $a$ the rate of bacterial division, $\alpha$ the rate of
attractant production, and $D_c$ the chemical
diffusion constant.  The terms in equation \ref{ks1} include the
diffusion of bacteria, chemotactactic drift and division of bacteria.
Equation \ref{ks2} expresses the diffusion and production of
attractant.

Equations of this type were first used to describe bacteria by Keller
and Segal
\cite{kel70}, and, with variations, have been the subject of extensive
investigations (see, e.g. \cite{murr89,ost89}). For {\it E. coli},
Schnitzer {\it et. al.} established the connection
\cite{sch90,sch93} between the time averaged properties of the
bacterial response and parameters in equations (\ref{ks1}, \ref{ks2}).
Thus, the extensive studies of individual bacteria provide a rigorous
justification for the equations, as well as measurements of the
coefficients.  There is one complication to this statement that is
worthwhile to mention: in 1975 Spudich and Koshland \cite{spud75}
showed that {\it E. coli} have ``non-genetic individuality'', manifest
in a distribution of tumble times (by about a factor of 2) between
genetically identical bacteria.  The consequence of this is that both
time-averaged and ensemble-averaged properties of the bacteria are
necessary to predict hydrodynamic coefficients; in addition, the
dynamics must be such that the ``distribution'' of bacteria in the
ensemble does not change with time.

It is convenient to nondimensionalize equations (\ref{ks1}, \ref{ks2})
by choosing a characteristic density equal to the maximum initial
density $ \rho_{ o} $.  The characteristic scale of attractant is
$D_b/k$.  The density then determines the length scale and timescale
according to $H = D_b D_c/( \alpha k \rho_{ o}) $ and $ t_{ o} = D_c/(
\alpha k \rho_{ o}) $.  Typical numerical values are $D_{ b} = 7
\mbox{x} \: 10^{ -6} \mbox{cm}^2$/sec; $D_{ c} = 10^{ -5}
\mbox{cm}^2$/sec; $k = 10^{ -16}
\mbox{cm}^5$/sec; and $ \alpha = 10^{ 3}
$/second/bacteria.  For an experiment \cite{bre98} which has $ \rho_{
o} = 10^{ 6} /\mbox{ cm} ^ 3 $, the length scale is 260 microns and
the timescale 100 seconds.  The equations become
\begin{eqnarray}
\frac{\partial \rho}{\partial t} &=& \nabla^2 \rho - \nabla\cdot(\rho\nabla c) 
+ \delta \rho \label{density}\\
\epsilon \:  \frac{\partial c}{\partial t} &=& \nabla^2 c + \rho\label{attract},
\end{eqnarray}
where $\epsilon=D_b/D_c$ and $\delta=a t_o$. For the experiments shown
in Figure 1, cells divide much more slowly than the dynamics take
place: the timescale for cell division is $\approx 2 $ hours, giving $
\delta \approx 0.01 $. Therefore we approximate $\delta =0$.  The value of the
parameter $ \epsilon $ varies: for experiments in semi-solid agar, the
diffusion of bacteria is much slower than attractant diffusion, which
motivates the limit $\epsilon=0$ \cite{bre98}.  For experiments on
bacteria in a liquid culture, $\epsilon \approx 1$.  We will consider
both of these limits in this paper.  The $ \epsilon = 0 $ limit is
particularly convenient for asymptotic calculations; we will use
$\epsilon=0$ when doing analytic calculations. Our numerical
simulations give results independent of $\epsilon$ in the range
between 0 and 1.

For analytic calculations, working with the mass is particularly
convenient, as will become apparent below. For reference, we show the
form of the equations here. Define the mass contained within a
radius $r$ as
\[
m(r) = \int dr \: r^{d-1} \rho.
\]
This definition
(and the limit $\epsilon=0$) allows us to eliminate the concentration
and write the original equations as
\begin{equation}
\frac{\partial m}{\partial t} = r^{d-1} \frac{\partial \rho}
{\partial r} +  \rho m	\label{mass}
\end{equation}

\section{Cylindrical Collapse}

In this section we construct the solution for a uniformly collapsing
cylinder of bacteria, according to the evolution equations (3,4).
Although the existence of cylindrical chemotactic collapse is well
known \cite{chi84}, the asymptotic solution describing the collapse
has never been constructed. Herrero and Velazquez \cite{her96}
understood important features of two-dimensional collapse and
attempted to find the solution; however, their solution is different
from what we observe in numerical simulations (see Figure \ref{length}) and is therefore
probably unstable. 

The usual derivation of a similarity solution derives  a set of ordinary differential equations  from the scaling laws suggested by dimensional
analysis. As shown below, the major
difficulty arises because in the present situation these ODEs
have no solutions consistent with the boundary conditions.
This breakdown of  ``dimensional'' scaling only occurs in two-dimensional collapse.  In higher
dimensions, the scaling laws suggested by dimensional analysis work
fine \cite{bre97}.  The overall 
structure of this problem is similar to critical
collapse in the focusing nonlinear Schrodinger equation
\cite{pap99}, where the initial proposal for constructing the
asymptotic solution involved rather intricate mathematics (continuation
as a function of spatial dimension).  Recently,
a more physical derivation was put forward \cite{pap99}.
  This work on the nonlinear Schrodinger equation inspired our
construction of the solution for chemotactic collapse.  Our central
result is that the density singularity shows corrections to the
dimensional scaling laws.  Defining $ \tau = t^{*} -t $ the distance
to the singular time, the dimensional scaling law is $\rho =
\tau^{-1}$. In the transient regime, we find a slow
power law correction to this dimensional scaling law, $ \rho \approx
\tau^{-11/8}$. After the transient asymptotic regime,  logarithmic corrections
are present, of the form
\[
 \rho =\frac{| \log |\log \tau||^{-1/2}}{ \tau }.
\]
Although the transient regime is present for 10 decades in the
collapse time, we believe that this {\rm log log} law is the
asymptotically correct blow up rate.  A {\rm log log} correction also
appears for 2D critical collapse in the nonlinear Schrodinger equation
(See
\cite{pap99} and the references therein).

\subsection{Critical Dimension for Collapse}

The competition between dissipation and collapse leads to a critical
dimension. In this problem, the critical dimension  is 2, and one
dimensional collapse---that is, collapse to a planar structure 
with infinite density---is forbidden.

We make qualitative arguments to explain the critical dimension by
comparing the chemotactic and diffusive fluxes in a contracting
structure.  First, no singularities exist for one-dimensional
contraction.  For a sheet of thickness $ \ell $ the diffusive flux is
of order (see equations (\ref{ks1}, \ref{ks2}))
\begin{equation}
J_{ \mbox{{\tiny D}} } \sim -D_b\frac{\rho}{\ell}.
\end{equation}
The chemotactic flux follows by integrating $D_c c'' \sim
\alpha\rho$ and defining $M^{ 1D}$ as the mass per unit area of the planar
region. Then the chemotactic flux is
\begin{equation}
J_{ \mbox{{\tiny C}}}\sim k \rho c' \sim \alpha k \rho M^{ 1D} D_c^{-1}.
\end{equation}
If the system collapses onto a plane, the thickness of the sheet
$\ell\to 0$. In this case the diffusive flux blows up while the
chemotactic flux is unchanged. Thus a planar region with small
thickness is unable to form infinite density, because diffusion
eventually stops the collapse.

The situation is different for higher dimensional structures.  For
symmetric spherical collapse (three directions contract
simultaneously), the chemotactic flux is singular. When we balance
$D_c \nabla^2 c \sim \alpha \rho$, we find $(r^2 c')' \sim \alpha r^2
\rho/D_c$. This implies a concentration gradient  $c' \sim \alpha M^{ 3D}/
(\ell^2 D_c)$, where $M^{ 3D}$ is the mass contained within a radius
$\ell$.  The net inward flux of bacteria is then
\[
J \sim \frac{-D_b \rho}{\ell} + \frac{\alpha k  \rho M^{ 3D} D_c^{-1}}{\ell^2}
\]
As $\ell \to 0$ the inward flux (second term) dominates and collapse
occurs.

In two dimensions we encounter a subtlety. Assuming cylindrical
collapse and repeating the dimensional argument, we have $(r c')' \sim
\alpha r\rho/D_c$, and $c' \sim \alpha M^{ 2D}/ (\ell D_c)$. (Here
$M^{ 2D}$ is the mass per unit length of the cylinder.)  The inward
flux is
\[
J= \frac{-D_b \rho +  \alpha k  \rho M^{ 2D}  D_c^{-1}}{\ell}
\]
Two dimensional collapse is critical: both fluxes have the same
scaling with $\ell$. According to this simplified argument, there is a
net inward flux if $M ^{ 2D}> D_c D_b/(\alpha k) $, which suggests
that a system with mass above this critical value collapses.

\subsection {Similarity Solutions}

We now quantify the preceding dimensional arguments and collect the
known solutions to the chemotactic equations. (As discussed above, the
analytic solutions are derived with $ \epsilon = 0 $.) First, consider
one-dimensional collapse.  Making the substitution\cite{bre98}
$v=\nabla c=\partial_x c$ in equations (\ref {density}, \ref{attract}) implies
\[
\frac{\partial v}{\partial t} = \frac{\partial^2 v}{\partial x^2}
- v \frac{\partial v}{\partial x}
\]
This equation is the Burgers' equation; singular solutions
to this equation do not exist \cite{whitham}.

In $d=2$ and higher, density singularities can develop. In three
dimensions, the nature of the blowup is straightforward.  The characteristic
length scale ($L $) varies in time, and the spatial structure is determined
by the changes in $L $.  A singularity corresponds to $L\rightarrow 0
$. We guess the form of the similarity solution by balancing the
different terms in equations (\ref{density},\ref{attract}).  The
diffusive dynamics imply $L=\sqrt{t^*-t}=\sqrt{\tau}$, with $t^*$ the
singular time and $\tau$ the distance to the singular time. Defining a
dimensionless similarity variable $
\eta = r/L $, we find the scaling form of the density, concentration,
and mass:
\begin{eqnarray*}
\rho &=&\frac{1}{L^2} R (\eta) \\
c & = & C ( \eta)	\\
m &=& L^{d-2} M (\eta)	
\end{eqnarray*} 
In writing this form of solution, we have assumed radial collapse at
the origin ($r=0)$.  For a similarity solution to be valid, it must
obey the correct boundary conditions: the density $\rho$ and the
attractant concentration $c$ must be time-independent far from the
singularity, which requires $R \sim \eta^{-2}$ and $C \sim$ constant
as $\eta \to \infty$.

Plugging in the scaling form gives an
ordinary differential equation in the similarity variable $\eta$. (Throughout this discussion, $d $ refers to the number of simultaneously contracting dimensions.)
\begin{eqnarray}
\frac{ \eta M'}{2} &=& \eta^{d-1} R' + RM \\		\label{sim}
\eta^{d-1} R &=& M' .					\label{sim2}
 \end{eqnarray}
In $d=3$, the similarity equation can be 
solved exactly; the one stable solution, found by Kadanoff \cite{bre97}, is
\begin{equation}
R=\frac{4(3+\eta^2)}{(1+\eta^2)^2}
\end{equation}
As demonstrated in \cite{bre97}, this solution well describes
numerical solutions.

\begin{figure}[p]
\centerline{\epsfysize=3in\epsfbox{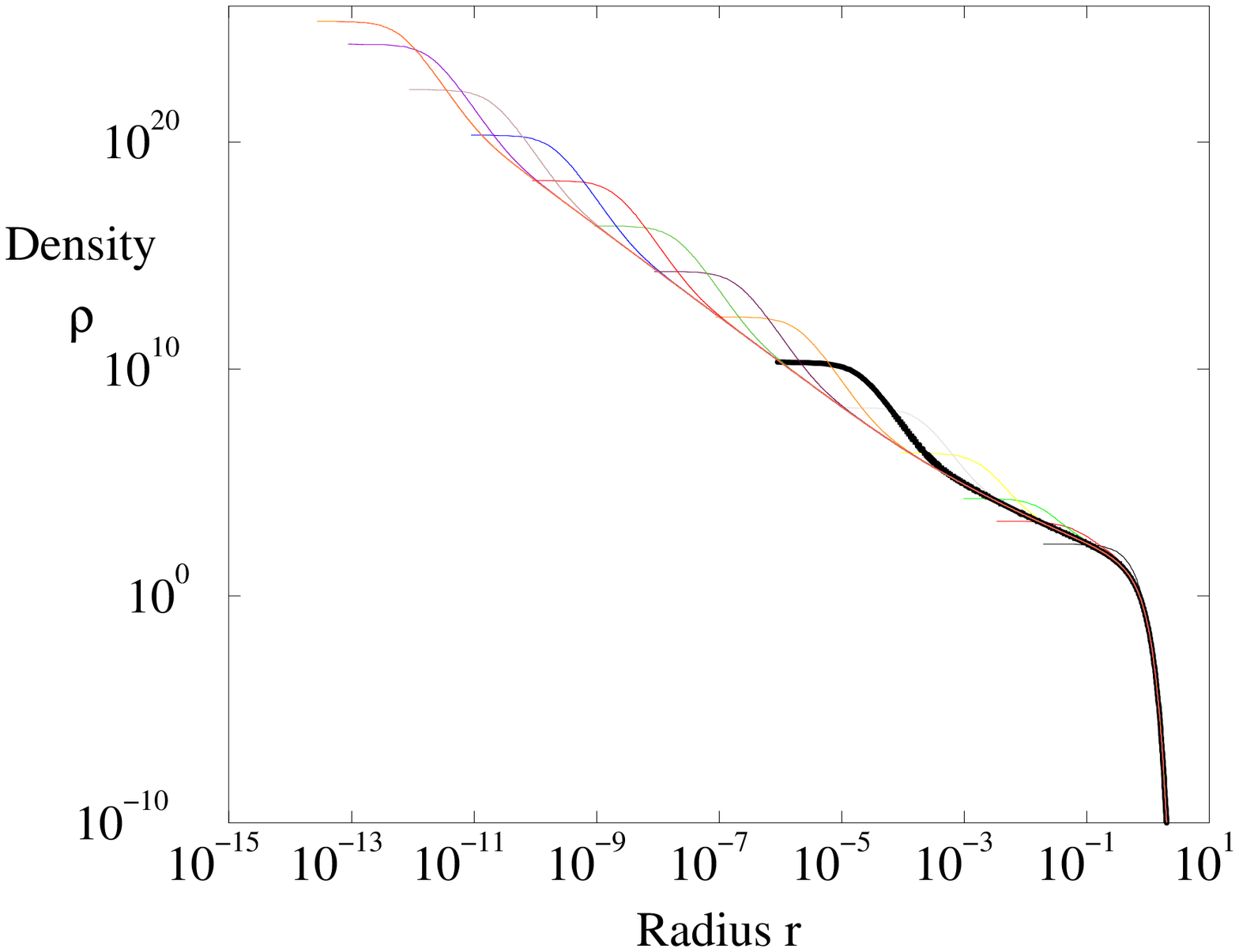}
\epsfysize=3in\epsfbox{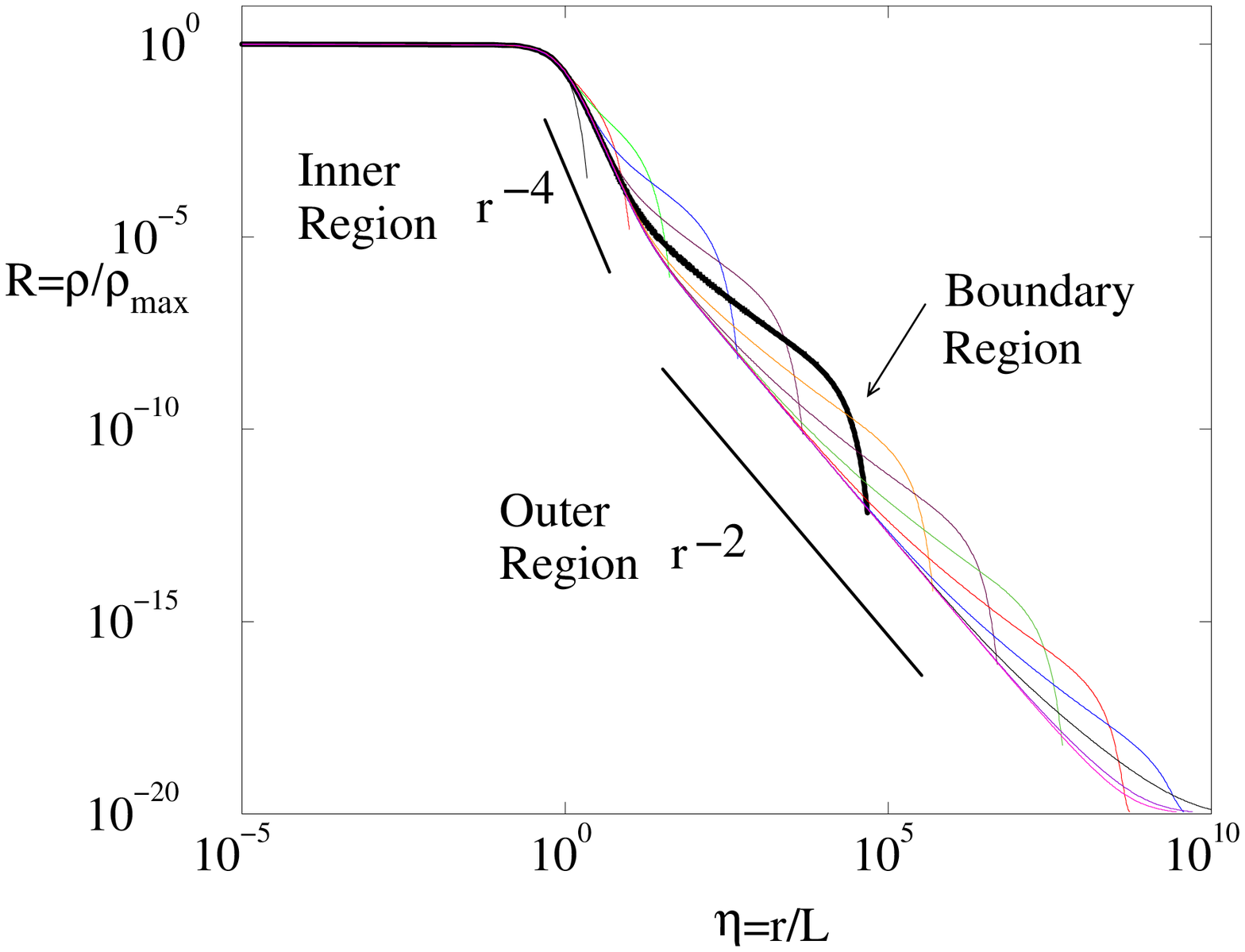}} 
\caption[]
{Density profiles for a collapsing cylinder. Different lines
correspond to different times approaching the singularity. On the
right, the density has been rescaled by the maximum density and the
radius rescaled by the width.  The solid line shows a profile at the same time both in this Figure and in Figure \ref{cmeas}. The crossover between the stationary
solution, for which the large $\eta$ behavior scales as $\eta^{-4}$ and the
true outer solution, which scales is $\eta^{-2}$, is apparent.  }
\label{logstat}
\end{figure}

\subsection{Solution in two dimensions}

In two dimensions, there is no solution to equations (\ref{sim},\ref{sim2}) which
satisfies the boundary conditions. To see this, note that the
similarity equations can be integrated to give
\begin {equation}
R = e^{ \frac{\eta^{ 2}}{4}} e^{\int\frac{M}{ \eta}}
\end {equation}
This form for $R $ cannot satisfy the boundary conditions that the
density and mass be stationary at large $\eta$, because $R $ grows
without bound as $ \eta \rightarrow \infty.$

Nevertheless, a similarity-type solution to the equations exists, as
Figure \ref{logstat} shows. The density profiles at different times
(left) have been collapsed by rescaling the length scale (right). This
figure shows the basic features of the solution: in the ``inner''
region, there is a scaling solution. This matches onto an ``outer''
region far from the singularity, with a final ``boundary'' region
where the boundary conditions must be satisfied.  We illustrate this section with plots from a single numerical simulation.  The numerical method is described in Appendix B; its most
important feature  is the remeshing, which moves
mesh points every time the maximum density increases by one percent to
better resolve the singularity \cite{simulation1}.

\begin{figure}[p]
\centerline{\epsfysize=3in\epsfbox{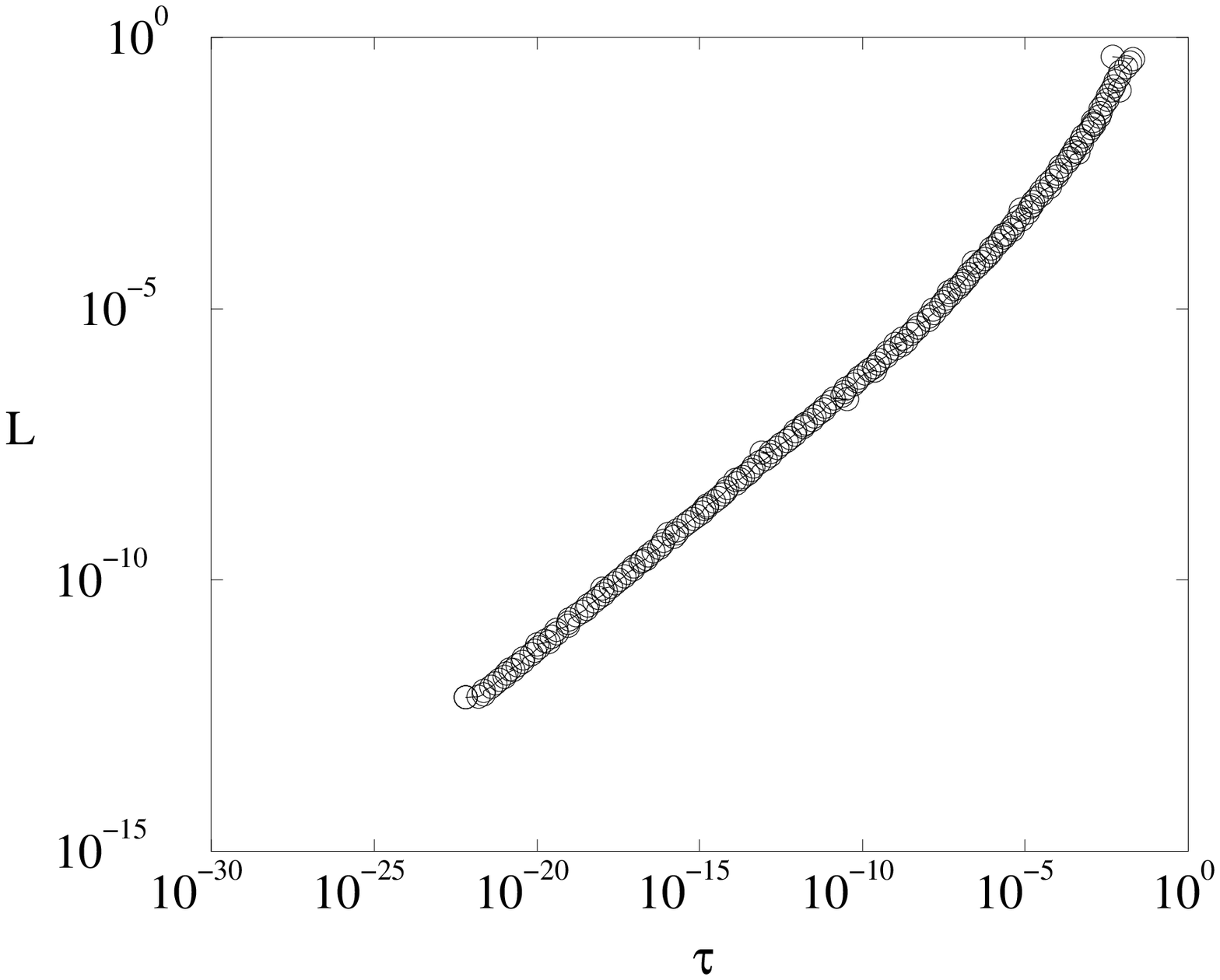}\epsfysize=3in\epsfbox{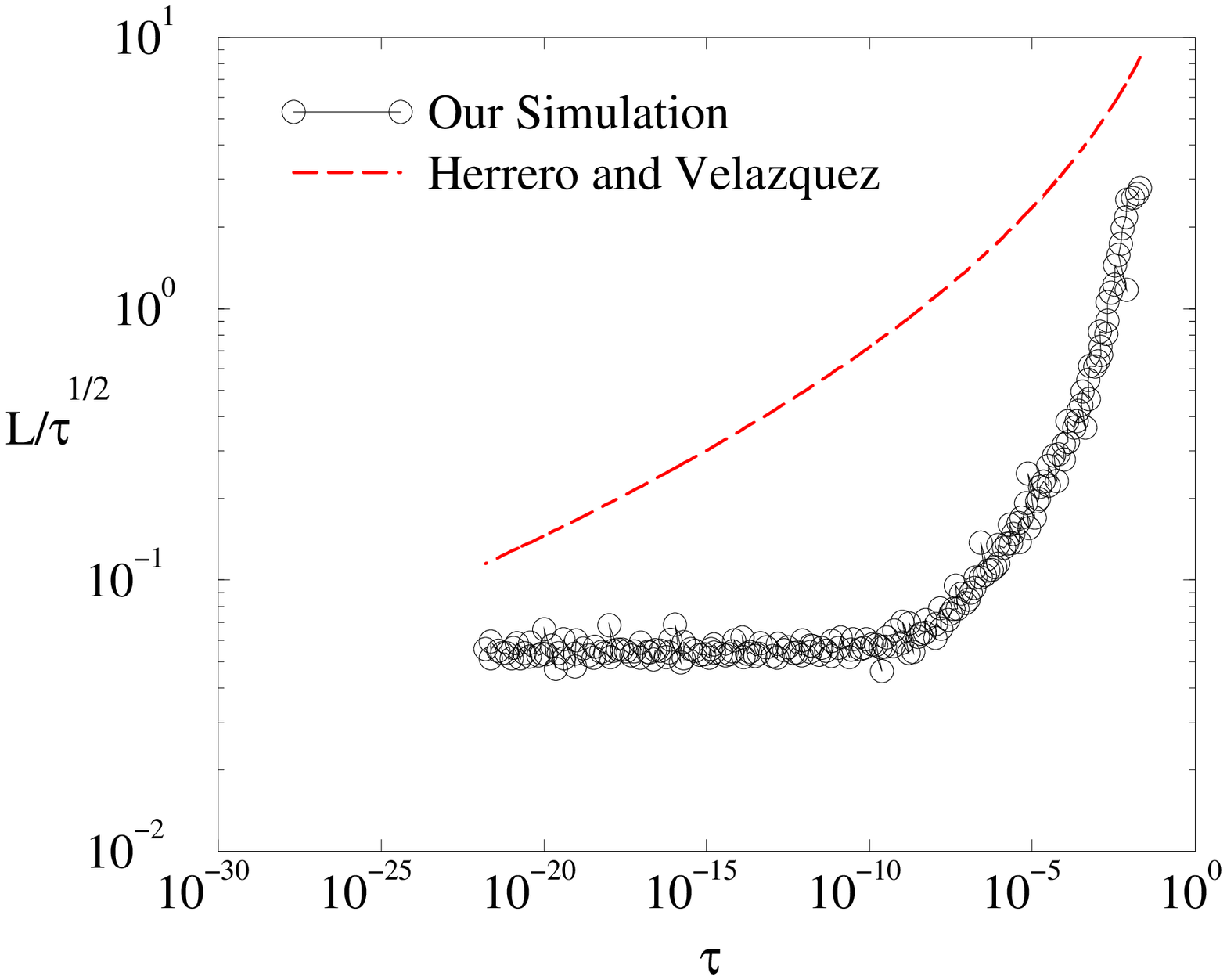}}
\caption[]
{Time dependence of the length scale, $L$. In simulations, the length scale is
defined as the length over which the density decreases by a factor of
five, a characteristic of the
inner region. The first plot shows $L$ versus distance to the
singular time, which obeys close to a square root law. The second plot
shows $L/\sqrt{\tau}$ versus $\tau$, showing evidence of the transient
regime ($L/\sqrt{\tau} \sim \tau^{3/16})$ and, closer to the
singularity, the asymptotic regime with logarithmic corrections to the
length scale.  For comparison, we have plotted the predictions of
Herrero and Velazquez \cite{her96}.
\label{length}}
\end{figure}

How do we reconcile the numerical (and experimental) observations of
two-dimensional collapse with the argument above that no solutions
exist? We argue here that corrections to the similarity solution arise
to solve this problem.  We
find in numerical simulations that although the basic self-similar scaling
$\rho_{m} \sim L^{-2}$ is preserved, the time dependence of $L$ is
different than the simple dimensional argument suggests.  Figure
\ref{length} shows the scaling of $L$ with time: if the
dimensional law $L\sim \sqrt{\tau}$ held, then $L/\sqrt{\tau}$
(plotted on the right) would be constant. Our simulations indicate a more complex time dependence than $L\sim \sqrt{\tau}$.  We argue that the problem develops two timescales: a fast timescale
on which collapse happens, and a slow timescale corresponding to
changes in the similarity profile. Working in ``similarity variables''
allows a separation of these two timescales: we eliminate the known,
fast time dependence so we can analyze the slow evolution of the
system.

\begin{figure}[p]
\centerline{\epsfysize=2.5in\epsfbox{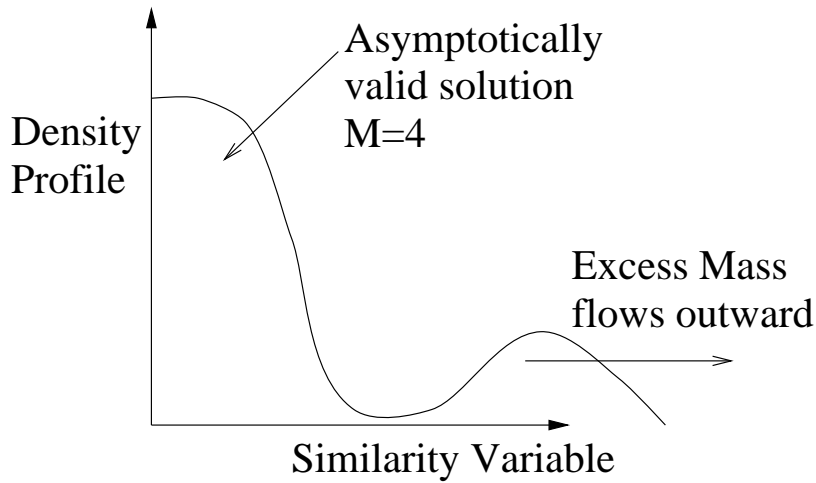}}
\caption[]
{Schematic of cylindrical collapse.
\label{logsketch}}
\end{figure}

As $L\to 0$, the inner collapsing region converges to a similarity solution which we call the {\sl stationary}
solution because it has the same form as the stationary solution to
the original equations.  This asymptotic solution has a dimensionless
mass of precisely 4.    Thus, the evolution of the similarity solution has a
specific physical interpretation.  In dimensionless (similarity)
variables, an inner region expels any excess mass to approach $M=4$.
(Figure \ref{logsketch}).  Matching between the inner and outer regions determines the dynamics.

\subsubsection{Separation of Time Scales and Inner Solution}

To find the solution, we use techniques originally applied to a
similar problem in the 2D nonlinear Schrodinger
equation\cite{pap99}. We first define an important quantity: the
collapse rate $A(t)=-\dot{L} L $.  (Note that the collapse rate of the
system is $\dot{L}/L $; we can make a dimensionless collapse rate by
multiplying by the timescale $ \tau \sim L^2$.  Thus $A$ is the
collapse rate.)  For ``regular'' collapse, the collapse rate
$-\dot{L}{L}=1/2$ is a constant. In the presence of corrections, the
collapse rate goes asymptotically to zero.

We rewrite the similarity equation assuming that both quantities $M $
and $R $ depend on the similarity variable $ \eta $ and (slowly) on
time.  The similarity equation is then
\begin{eqnarray}
L^2 \frac{\partial  M}{\partial t} + A \eta M' &=& \eta R' + RM
\label{masseqn} \\
\eta R &=& M' .\label{masseqn2}
\end{eqnarray} 
Note that here the time derivative of $M$ refers 
only to {\em explicit} time dependence of the mass; the second term in 
the equation takes into account the time dependence slaved to the
varying length scale.

\begin{figure}[p]
\centerline{\epsfysize=3in\epsfbox{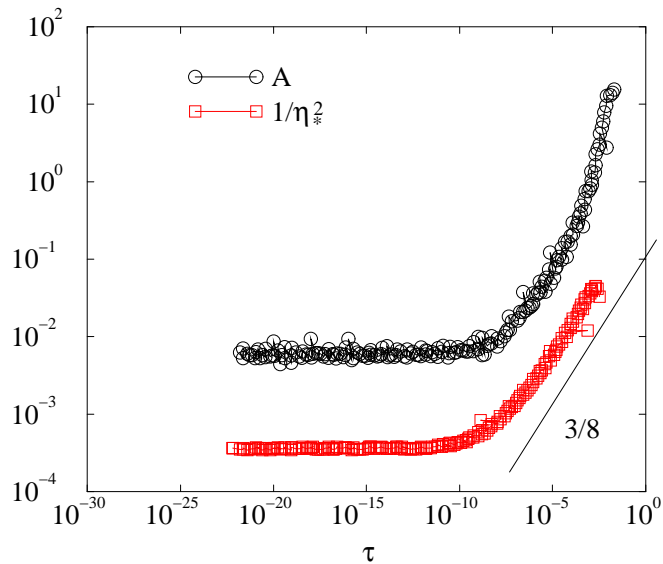}}
\caption[]
{Plot of the collapse rate $A$ and $\eta_*^{-2}$ versus
distance to the singular time $\tau$. The line shows a power law $A
\sim \tau^{3/8}$. $A$ is measured by evaluating $\dot{L} L$
numerically; $\dot{L}$ is the difference in length scale between two
time steps, divided by the time step size. The length scale $L$ is
defined as the radial coordinate where the density decreases by a
factor of 5. Similarly, $\eta_*$ is measured as the coordinate where
the numerical density profile differs from the exact stationary
solution (Equation (\ref{statden})) by a factor of 4.}
\label{ldotl}
\end{figure}

We can solve the similarity equation in the inner region.  As the
collapse proceeds, it does indeed slow down.  In fact, a numerical
measurement of $A$ (Figure
\ref{ldotl}) shows that the collapse rate decreases by 3 orders of
magnitude during the initial stages of the collapse. This motivates us 
to look for a solution with $A=0$; that is, a stationary
solution (in similarity variables). This stationary
solution solves Equations (\ref{masseqn},\ref{masseqn2}) when $A=0$ and $ \partial_{ t} M=0$.
The equation for the mass is then
\[
 \eta M'' + M' (M-1) = 0.
\]
Exact analytic solutions to this equation are
\begin{eqnarray}
R_o &=& \frac{8 }{(1+\eta^2)^2}	\label{statden}\\
M_o &=& \frac{4 \eta^2}{1+\eta^2}		\label{statmass}
\end{eqnarray} 

From the formula for $M_o$, we see an important feature of the
stationary solution: the total mass is 4 in dimensionless units. This
reflects a rigorous result\cite{chi84,bre97}: collapse will occur if and
only if the total mass per unit length of the cylinder satisfies $M>
4$.  For $M<4 $, no collapse is possible---the system evolves to
constant density.  For $ M > 4 $, collapse occurs; however, numerical
simulations show that the collapse converges toward a solution with a
collapsing mass precisely equal to four.  In similarity variables,
therefore, mass flows {\sl away} from the origin.

Compare this expression for $R_o$ to the numerical density profiles in
Figure \ref{logstat}. The shape of the profile is accurately described
by the formula for $R_o$, which confirms that the stationary solution
holds in the inner region. At large $\eta $, the stationary solution
has $R_o \sim 8
\eta^{-4} $, while the boundary conditions require $R \sim c
\eta^{-2}$.  This ``inner'' solution to the equations must therefore
match onto an ``outer'' solution, as shown in Figure \ref{logstat}.
How to analyze this matching is the subject of the rest of this
section.

Where does the crossover between inner and outer region occur? We can
estimate the location of the crossover by noting that the crossover
between inner and outer solutions will happen at some coordinate
$\eta_*$, when $A\eta M' \sim \eta R' $ (see Equation (\ref{masseqn}))
This gives
\begin{equation}
\eta_* \sim A^{-1/2}.
\label{etaA}
\end{equation}
The proportionality between $A $ and $ \eta^{ -2}_{ *} $ is clear in
numerics (Figure \ref{ldotl}).

\subsubsection{Transient Regime}

Near the beginning of the collapse, the singular (inner) part of the
solution has formed, but still strongly feels the influence of the
boundary.  During this initial, transient regime in the dynamics,
excess mass is expelled. At the end of the transient
regime, the true asymptotic state is reached. The transient regime is
clear in the simulation shown here (Figures
\ref{length}, \ref{ldotl}, \ref{rho}), where it persists until $\tau \approx
10^{-10}$. Here we attempt to analyze the transient regime and
consider how the system approaches the asymptotic regime.  This
analysis is difficult, because the transient dynamics depend on the
initial and boundary conditions of the system, and hence is
particularly difficult to characterize analytically.

\begin{figure}[p]
\centerline{\epsfysize=3in\epsfbox{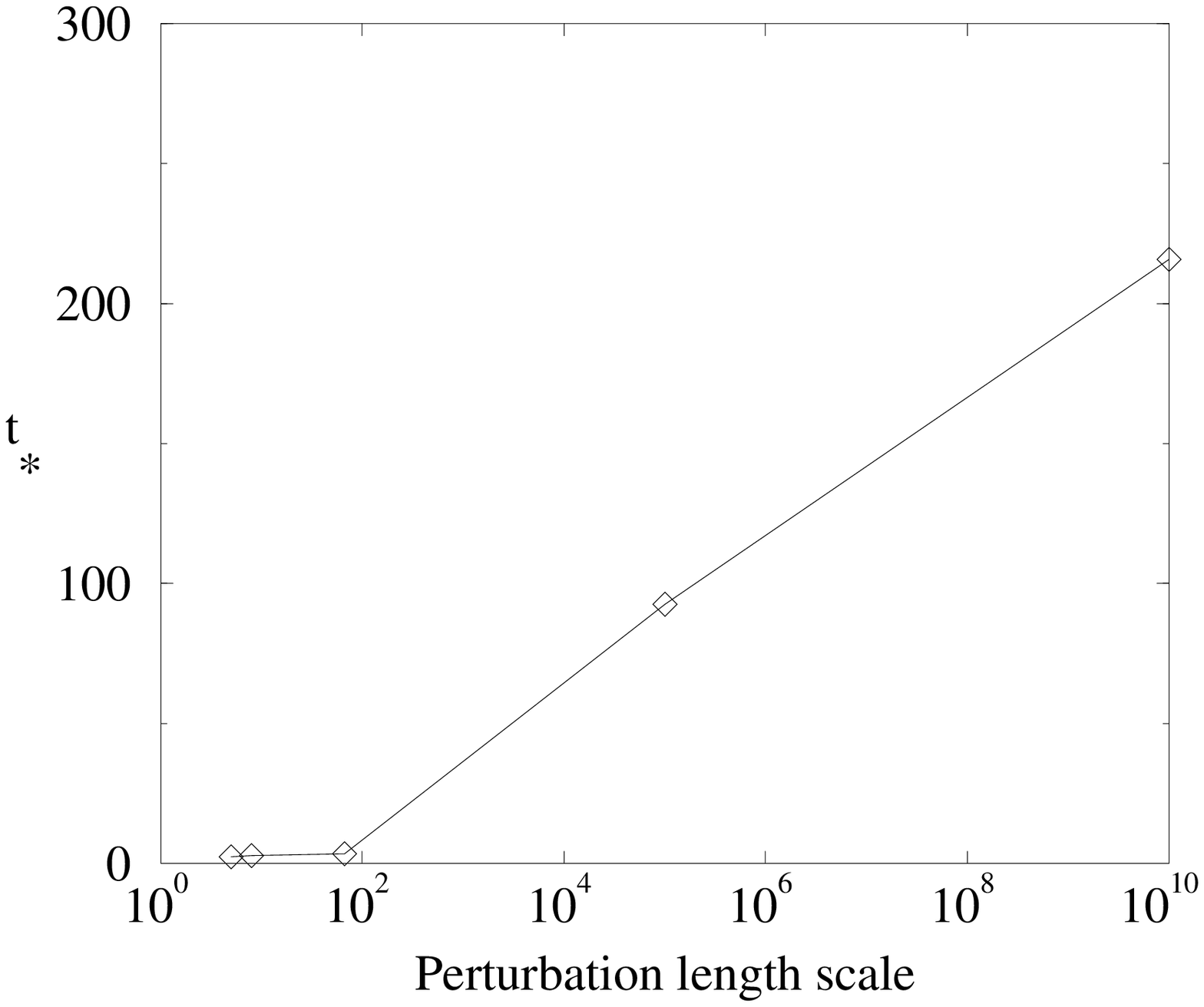}\epsfysize=3in\epsfbox{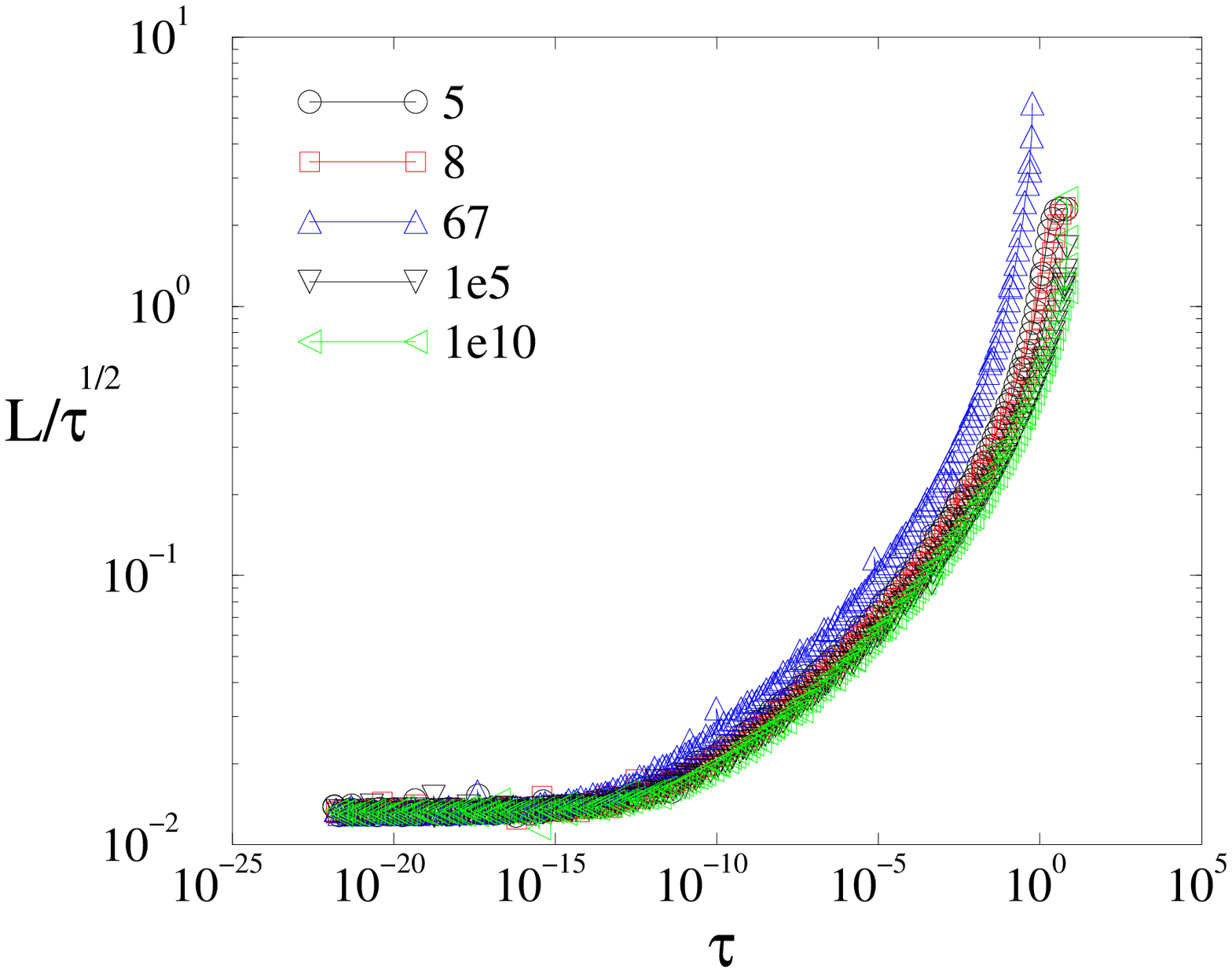}}
\caption[]
{Comparison of different initial conditions. On the left, the time
until the singularity is reached depends logarithmically on the length
scale of initial density variation, as expected for an unstable
collapsing mode. On the right, the transient regime dynamics are
almost identical, despite the wide range in length scale of
perturbation. The legend shows the length scale of initial
density variation for each curve.  }
\label{initcond}
\end{figure}

Why even bother with the transient behavior?  Because, as we mentioned
above, the transient regime persists until $\tau \approx 10^{-10}$.
The density increases by 10 orders of magnitude before the true
asymptotic dynamics are reached\cite{transient}.  Therefore, all
experiments can probe only the transient regime.  Furthermore, the
detailed dynamics of the transient regime are not highly sensitive to
the initial conditions. We did a number of simulations in which the
length scale of the initial density distribution was varied (Figure
\ref{initcond}). When the density is initially almost constant, the
time at which the singularity is reached, $t_*$, is large compared to
a simulation in which the density is initially peaked. However, the
transient dynamics are almost identical, regardless of whether the
density is initially constant, or highly peaked \cite{simulation2}.

Our discussion of the transient regime that follows is incomplete and
{\it ad hoc}.  It is, however, the only self consistent explanation of
numerical results we were able to find. We welcome additional, more
conclusive work on this question. 

We argue
that the corrections to the length scale in the transient regime are
faster than logarithmic.  Although
theoretically we cannot rule out logarithms,  we
were not able to find an asymptotic solution for a correction which
depended purely logaritmically in time that explained all features of
the numerical simulations.  For this reason we believe that the most
consistent interpretation of the transient regime is that it actually
breaks the scaling laws predicted by dimensional analysis and
introduces different exponents. 

Here we use
features of the numerical solution to guide our construction of the
transient correction; this argument, while not entirely
satisfactory, is at least consistent with all the simulations.  We
seek corrections to dimensional scaling of power law form: that is, $L
\sim
\sqrt{ \tau} \tau^{ \beta} $ and we are determining $\beta$.
As long as $\beta > 0 $ , we still have asymptotic validity of the
analysis, because $A $ goes to 0.  (Note that $ \beta $ depends on the
initial and boundary conditions, and is not universal.)

\begin{figure}[p]
\centerline{\epsfysize=3in\epsfbox{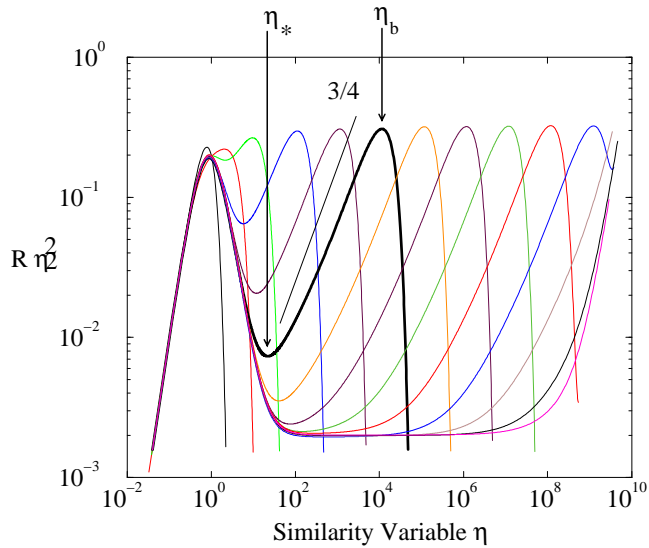}}
\caption[]
{A plot of $R \eta^2$ versus $\eta$, used to construct the corrections
in the transient regime. Profiles are for different times. The line
shows a power law $\eta^2 R \sim \eta^{3/4}.$ The heavy line shows the
profile at the same time as the heavy line in Figure \ref{logstat}.  }
\label{cmeas}
\end{figure}

\begin{figure}[p]
\centerline{\epsfysize=3in\epsfbox{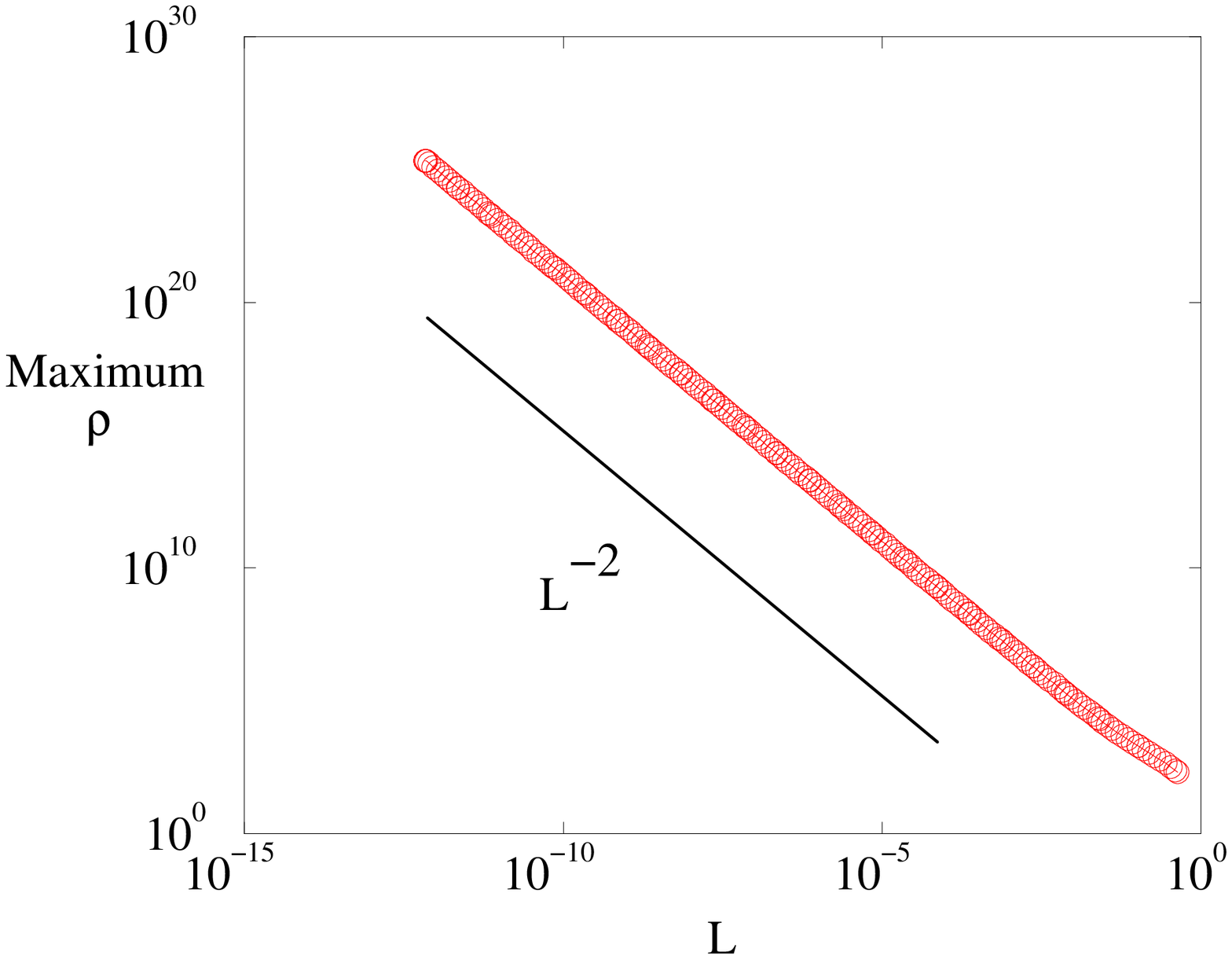}
\epsfysize=3in\epsfbox{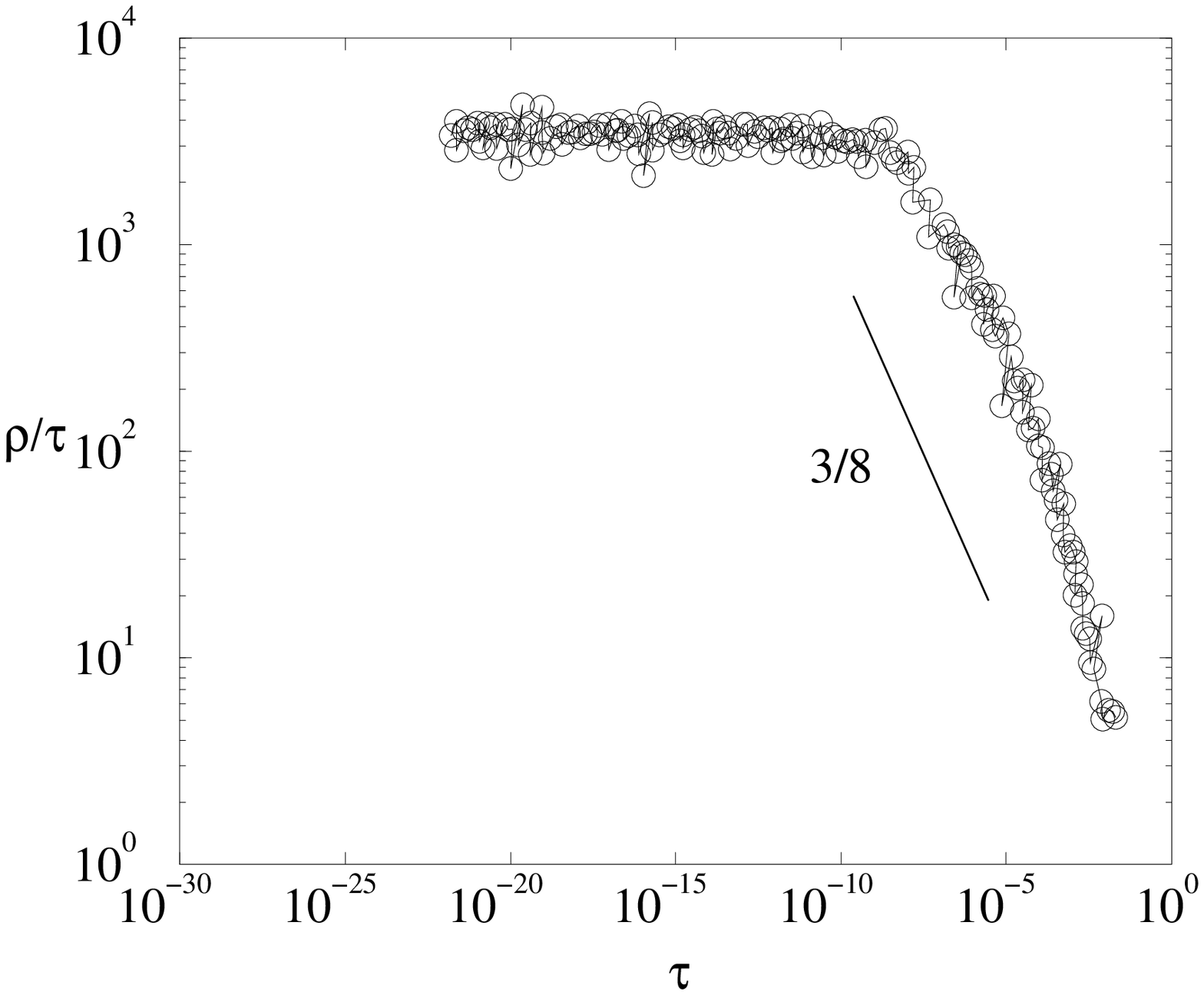}} 
\caption[]
{The scaling of the maximum density with time. On the left, $\rho_m$
versus $L$, which is well described by the expected scaling. On the
right, $\rho_m \tau$ versus $\tau$, showing the transient and
asymptotic regimes.} 
\label{rho}
\end{figure}

We can distinguish between the transient and
asymptotic regimes by looking at Figure \ref{cmeas}. This plot shows
profiles of the normalized density ($R(0)=1$) times $\eta^2$. At later
times, the curve shows a flat region, demonstrating that $R \sim
\eta^{-2}$ over a large region. This is the hallmark of the true
asymptotic regime: the stationary solution is valid in the collapsing
region, then $\eta^{-2}$ outer region exists far from the singularity,
which in turn connects to the boundary. For earlier times, however,
the flattened region is not present: instead the forming singularity
connects directly to the outer ``bump'' where the excess mass has been
pushed. \cite{boxsize}

The inner solution must match on to the
``bump'' that makes up the outer region \cite{boundarycondition}. The
bump corresponds to 
the excess mass in the system, which is stationary in real
coordinates.  Thus the position of the bump in similarity variables is
$
\eta_{ b} \sim L^{-1} \sim \tau^{ -1/2- \beta} $, where we have assumed power law
corrections to the length scale.  By reading off the plot (Figure
\ref{cmeas}) we can see two important features of the outer region:
first, the value of $ \eta^{ 2} R $ is constant at the position of the
bump: $
\eta_{ b}^{ 2} R_{ b} = \gamma $.   Next, the form of the solution from
the crossover point to the bump ($ \eta_{ *} < \eta < \eta_{ b} $)
is approximately a power law, which we write
\[
\eta^{ 2} R = c \eta^{  \alpha} \mbox{ for } \eta_* < \eta < \eta_b
\]
The exponent $ \alpha $ is a free parameter that we find from the
simulations.  Taking $\alpha$ from the numerics (reading off Figure
\ref{cmeas}) uniquely determines all other scaling exponents and hence
gives a consistency check. To determine $\beta$, combining the
relations above lets us determine the time dependence $c \sim \eta_{
b}^{ - \alpha} \sim
\tau^{ \alpha (1/2 + \beta)} $.  This means that in the intermediate
regime $ \eta_{ *} < \eta < \eta_{ b} $ the solution obeys
\[
 \eta^{ 2} R \sim \tau^{  \alpha (1/2 + \beta)} \eta^{  \alpha}
\]
Finally, we match the two solutions at the crossover $ \eta_{ *} $,
leading to the result $ \eta_{ *} \sim \tau^{ - \alpha (1/2 +
\beta)/(2 + \alpha)}$. Demanding consistency in time scaling we must
have $\eta_{*}^{-2} \sim A \sim \tau^{2 \beta}$.  This determines $
\beta $:
\[
\beta=\frac{\alpha}{4}
\]
For the simulations shown, $\alpha=3/4$. We have examined $\alpha$ for
a variety of initial conditions and find it always to be close to this
value.  This one  assumption for $\alpha$ leads to 
predictions for other measureable exponents, all of which are in
good agreement with numerical simulations.

Thus in the transient regime we have
\begin{eqnarray*}
L &=& \sqrt{\tau} \tau^{3/16}	\\
A &=& \tau^{3/8}		\\
\rho_m &=& \frac{1}{\tau \tau^{3/8}} = \tau^{-11/8}
\end{eqnarray*}
These scalings compare well with numerics, shown in Figures
\ref{length},\ref{ldotl}, and \ref{rho}.

\subsubsection{Asymptotic Regime}

Once a true scale separation exists between the inner (collapsing)
region and the boundary, the matching is different. Here we see slow
(logarithmic) corrections in numerics, motivating us to seek a
logarithmic correction to the length scale. We define the slow time
scale $s=-\log \tau$, and seek a length scale $L$ of the form
$L=\sqrt{\tau}/f(s)$. With this modification, Equations (\ref{masseqn},\ref{masseqn2}) for
the mass becomes
\begin{equation}
\frac{1}{f(s)^2} \frac{\partial  M}{\partial s} + A \eta M' = \eta R' + RM
\label{logeqn}
\end{equation}
The inner region
must match onto an ``outer'' solution of the form $R=c \eta^{-2}$
The value of the coefficient $c$ can be estimated from the total mass
and size of the system. If $c$ is independent of time, the total
mass $M$ in the far field is logarithmically divergent: $M\sim c \log
\eta$. However, conservation of mass requires that the total mass be
constant at the boundary: $M(W)=M_T$, which implies $c \sim
M_T/ \log(W/L)$ (up to logarithmic corrections) or
\[
c \sim \frac{M_T}{\log W + s}.
\]
As explained above, the crossover between inner and outer solutions
occurs at $\eta_* \sim A^{ -1/2}$. At $ \eta_{ *} $ we require that
the mass flux $ \partial M/ \partial s $ is continuous; this condition
connects the inner and other solutions, and fixes the dynamics of $f
$.

From the similarity equation (\ref{logeqn}) and $R(\eta)=c \eta^{-2}$
we have the mass flux in the outer region
\begin{equation}
\frac{\partial M}{\partial s} = - A f^2 c + O( \eta^{-2}) \quad \eta \to \infty
\end{equation}
The mass flux as $\eta \to \infty$ is spatially uniform.  This must
match the mass expelled by the inner region, which we find from $M_o
\approx 4 -  \eta^{-2}$. Equating fluxes at $\eta_*$ gives $\partial
M_o/\partial s |_{\eta_*} = - A f^2 c$ or
\begin{equation}
\frac{ \partial A} { \partial s} =  A f^2 c.
\label{flux}
\end{equation}
We can convert this to an equation for $f$ by using the definition of
$A=-\dot{L}{L}=f^{-2}(1/2+f'/f) \approx  f^{-2}/2$, where a prime
denotes derivative with respect to $s$. Plugging in the value of $c$
from above, we have an equation of the form
\[
\frac{ -f'} {f^3 } = \frac{M_T}{\log W + s }
\]
This relation holds for large $s$, giving the scaling of $f$
with time:
\[
\frac{ -f'} {f^3 } \sim \frac{1}{s}
\]
which gives
\[
f \sim \frac{1}{\sqrt{ \log |\log \tau|}}
\]
The asymptotic solution for two-dimensional collapse is a very slow
($\log \log \tau$) correction to the scaling laws from dimensional
analysis. Our simulations cannot verify the functional form of $f$;
however, we certainly see a crossover to slow (or nonexistent)
corrections to the length scale. In Figures
\ref{length},\ref{ldotl}, and \ref{rho}, the slow correction is
visible for $\tau$ between $10^{-20}$ and $10^{-10}$. A blow-up of this 
slow correction show that it is approximately constant, but dominated
by numerical noise---our numerics aren't good enough to resolve this
correction. Note that a simple logarithmic correction (that is, $f$
proportional to $\log \tau$) is {\em not} consistent with our
simulation, because such a function would not be approximately
constant over 10 orders of magnitude in time. Therefore, our numerics
are consistent with, but not proof of, a log log correction.


\section{Evolution of a Modulated Cylinder}

A collapsing cylinder eventually breaks into spherical aggregates.
Here, we find the envelope equation which describes how modulations to
the cylinder evolve. The challenge is to describe a {\sl collapsing}
cylinder.  Collapse amplifies initially small perturbations; therefore
small variations along the cylinder (in density and the radial length
scale) become large. We can perform a valid perturbation analysis by
studying variations in the singular time $t^*$.

In the original similarity solution, the singular time $t^*$ is
undetermined: if $t^*$ changes by a constant, the solution remains
valid. Allowing slow spatial variation in $t^*$ breaks this
symmetry. Therefore we expect the variation of $t^*$ to produce slow
dynamics in space and time.  Because this mode is the most slowly
decaying, it dominates the evolution of a cylinder. We derive a phase
equation, an approach used in many problems when stability is governed
by a slow mode associated with a broken symmetry\cite{manneville}.
Phase equations \cite{cross} were invented to understand problems like
convection, where the relevant symmetry is translation, and the
stability analysis is relative to a travelling wave solution. Earlier
research moved towards applying phase equations to singularities: In previous
work on blowup in the semilinear heat equation, Keller and Lowengrub
\cite{kel93} derived a transformation from a blowing-up variable to
one that vanishes; they then perturbed in the vanishing
variable. Also in the context of the semilinear heat equation, 
Bernoff \cite{ber99} has looked at how the singular time
varies along a cylinder.

We compare solutions of the phase equation to full numerics and show
that the evolution of a modulated, collapsing cylinder is well
described by the simplified phase equation. 

\subsection{Derivation of a Phase Equation}
 This section gives a flavor of the derivation of the phase equation,
 for details see Appendix \ref{phasedet}. We seek an equation for the
 dynamics of $\tau(z,t)=t^*-t= t_o + T(z,t)-t$. Slow variation
 requires that both $T''\ll T'$ and $\dot{T} \ll 1$. Once the collapse
 time varies along the axis of the cylinder, $R$ and $C$ are no longer
 exact solutions to the equations (\ref{density}, \ref{attract}).  We
 write the correction as
\begin{eqnarray*}
R = R_o(\eta) +  \delta  R_1(\eta, z,t) \\
C = C_o(\eta) +  \delta  C_1(\eta, z,t) 
\end{eqnarray*}
where $\eta=r/L_r $, with the radial length scale $L_r =
\sqrt{\tau(z,t)}/f(\tau)$, where $f$ is the correction to the length scale. The perturbation parameter $ \delta $ is of
order $L_r/L_z$, where $L_z$ is the scale of the density variation
along the axis of the cylinder.

We insert this guess into the original equations, and expand in $
\delta $.  The lowest-order equation gives the original similarity
equations.  At first order, we find an equation of the form:
\begin{equation}
\Lambda (R_1,C_1) = F(R_o,C_o) (\tau_t+1) + G(R_o,C_o) \tau_{zz} +
H(R_o,C_o) \frac{\tau_z^2}{\tau}
\label{phaseco}
\end{equation}
On the left side is a linear operator $\Lambda$ acting on $R_1$ and
$C_1$; $\Lambda$ comes from the linearization of the original
equations. The right hand side contains derivatives of the singular
time multiplied by known functions of $R_o$ and $C_o $.

Although $R_1$ and $C_1$ are unknown, the right hand side is
constrained by a solvability condition: Any function which is
annihilated by the adjoint of $\Lambda$ must be orthogonal to the
right hand side.  That is, if $\Lambda^\dagger g=0$ for a nonzero $g$,
then the inner product \cite{innerproduct}
\[ 
\langle g, \Lambda (R_1,C_1) \rangle  = \langle \Lambda ^{\dagger}
g , (R_1,C_1) \rangle = 0.
\]
Hence, the right hand side of (\ref{phaseco}) is orthogonal to $g$.
In this case (see Appendix \ref{phasedet}) precisely one nonzero $g$
satisfies $\Lambda^\dagger=0$. Taking the inner products leads to a
phase equation of the form
\[
c_1 (\tau_t+1) + c_2 \tau_{zz} + c_3\frac{\tau_z^2}{\tau} = 0
\]
where the constants $c_1, c_2, c_3$ can be expressed in terms of the
known functions $g$, $F$, $G$, and $H$:
\begin{eqnarray*}
c_1 &=& \langle g, F(R_o,C_o) \rangle	\\
c_2 &=& \langle g, G(R_o,C_o) \rangle 	\\
c_3 &=& \langle g, H(R_o,C_o) \rangle 
\end{eqnarray*}
As shown in appendix \ref{phasedet}, for a cylinder disintegrating
into spheres the phase equation is
\begin{equation}
\frac{\tau_t+1}{f^2} =\tau_{zz}
-\frac{\tau_z^2}{\tau} 
\label{phaselog}
\end{equation}
The correction terms in this equation arise directly from the
corresponding 
corrections to the length scale in the similarity solution. In the subsequent
section, we show that this results in asymptotically
different scalings for the radial and axial length scales on the
collapsing cylinder. (Hence,  a
``point'' singularitity that forms on a collapsing cylinder does not
have the same collapse rate as spherical collapse.)  In the
absence of  corrections to dimensional scaling, the phase
equation is simply
\begin{equation}
\tau_t+1 =\tau_{zz} -\frac{\tau_z^2}{\tau} 
\label{phase}
\end{equation}

\subsection{Numerical Simulations of a Collapsing Cylinder}

Now we compare solutions to equation (\ref{phaselog}) with a fully nonlinear simulation
of a collapsing cylinder. We
 have found two different mechanisms by
which modulations of the cylinder can produce singularities: The first
is a ``point'' singularity, in which the density blows up at a point on
the cylinder; the second is a ``travelling'' singularity, which moves
along the cylinder axis with a diverging velocity (as the singularity
is reached.)  

The primary technical difficulty in simulating a collapsing cylinder
is developing a remeshing algorithm to closely approach the
singularity.  The meshing algorithm described here resolves density
singularities along the axis of the cylinder with essentially
arbitrary resolution.  The algorithm is based on a simple one-dimensional scheme, which  redistributes mesh points every
$50$ timesteps to  resolve the singularity.  The two-dimensional
algorithm   uses the one-dimensional remeshing scheme along
both $\hat{r}$ and $\hat{z}$ simultaneously; the two dimensional
equations are then solved by operator splitting.  Details of the
algorithm are summarized in Appendix \ref{numerics}.

A typical simulation \cite{simulation3} started with a $z$-independent
initial condition, which was allowed to progress until the maximum
density reached $10^4$.  At this point, the radial profile of the
collapse was well approximated by the (2-dimensional) collapsing
similarity solution constructed in the previous section.  We then
added a $z$-dependent perturbation to the density profile, with
amplitude much smaller than the ambient density.

\begin{figure}[p]
\centerline{\epsfysize=3in\epsfbox{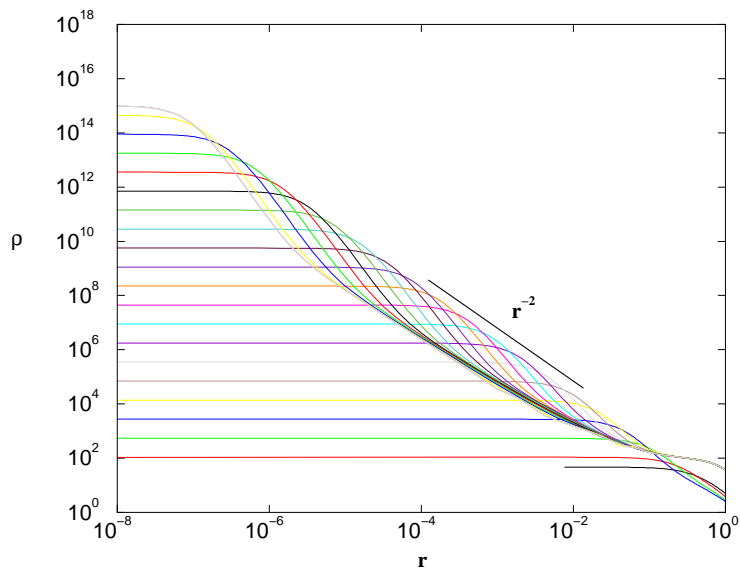}}
\caption[]
{Figure showing the {\sl radial} density profile at a single point on
a modulated, collapsing cylinder. Note the $r^{-4}$ region matched
onto the $r^{-2}$ region: The solution is well-described by the
two-dimensional similarity solution throughout the collapse
(cf. Figure \ref{logstat}) even in
the presence of a $z$ dependent modulation.}
\label{struct}
\end{figure}

The separation of scales hypothesis underlying the derivation of the
phase equation is maintained uniformly in time. We experimented with
different functional forms for the density perturbations; as long as
the length scale $L_z$ for variation in the $z$ direction is longer
than the variation in the radial direction $L_r$, perturbations tended
to grow. In all cases, the relation $L_z \gg L_r$ was maintained.  As
demonstrated in Figure \ref{struct}, a radial cross section of the
cylinder always revealed density profiles in agreement with the two
dimensional collapsing singular solution constructed in the previous
section.

Extensive simulations have found two types of density singularities
caused by modulations to the cylinder.

\subsubsection{Travelling Singularity}

\begin{figure}[p]
\centerline{\epsfysize=3in\epsfbox{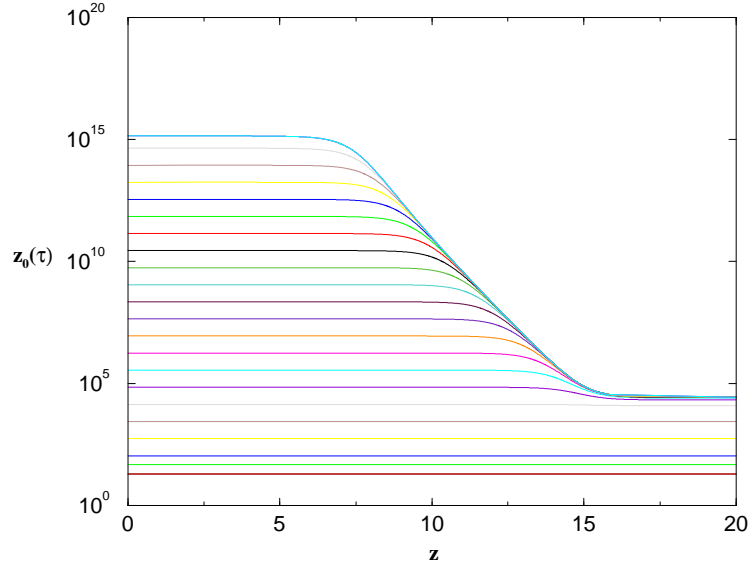}}
\caption[]
{Time evolution of the centerline density $\rho(r=0,z)$, with
a step function initial condition.  The 
perturbation is seeded when $\rho=10^4$, and subsequent 
profiles show when the maximum density increases by
a factor of five.  The highest density region {\sl propagates}
to the left.  Hence, the spatial extent of the high density
region shrinks as the collapse is approached.}
\label{prop1}
\end{figure}

{\sl Travelling singularities} occur when a step-like perturbation is
placed on the cylinder, increasing the density for $z>z_0$ and
decreasing the density for $z<z_0$.  The subsequent evolution occurs
at the boundary between these two regions. A simulation of this
process is shown in Figure \ref{prop1}.  The higher density region
propagates to the left.  Heuristically, the higher density region is
beginning to contract as a sphere, so its decrease in size is
consistent with the beginnings of spherical collapse.


This propagating singularity can be described as a solution
to the phase equation of the form
\[
\tau = \tau_o \: \phi(z - z_o(t))
\]
where $\tau_o=t^*-t$ is the basic phase expected from collapse. All
the nontrivial space and time dependence is absorbed in $\phi$ and
$z_o$.  Inserting into the phase equation (\ref{phaselog}), we have
\[
\frac{\dot{\tau_o} \phi - \dot{z_o} \tau_o \phi'
+1}{f^2} =\tau_o \left( \phi''-\frac{\phi'^2}{\phi} \right)
\] 
The right hand side of this equation is $O(\tau_0)$, which becomes
arbitrarily small as the singularity is approached. Therefore, the
left hand side must be equal to zero, which gives
\[\dot{\tau_o} \phi - \dot{z_o} \tau_o \phi'+1=0.\] 
If we demand the balance
$\dot{z}_0\tau_0= -A$, then 
$$z_0= A \log\tau_0.$$ 
The solution
for $\phi$ is then
\[
\phi(\eta)=1+e^{\eta/A}.
\]

Figure \ref{prop1} shows the density along the centerline of the
cylinder.
The decay of the highest density is exponential, as predicted by the
phase-equation solution constructed above.  A fit to the numerical
data shows that $$\rho(r=0,z) \sim e^{(-z/z_0)} \sim e^{(-3.1 z)}.$$

\begin{figure}[p]
\centerline{\epsfysize=3in\epsfbox{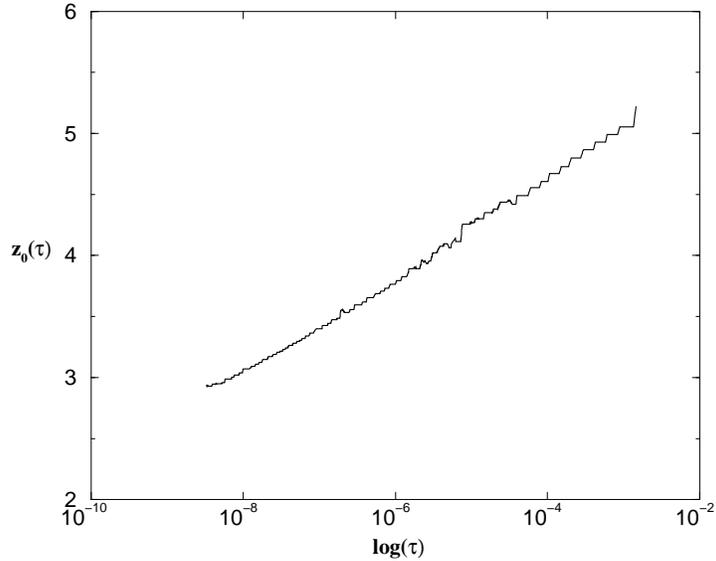}}
\caption[]
{Location of the edge of the maximum density region $z_0(\tau)$ as a
function of time.  As predicted by the theory, the edge moves
according to the law $z_0(\tau)= A \log\tau$.  Regression gives
$A\approx 0.17$.}
\label{xmax}
\end{figure}

Figure \ref{xmax} shows the location of the edge of the maximum
density region $z_0(\tau) $ as a function of $\tau$.  As predicted by
the phase equation analysis, the figure shows that $z_0= A \log\tau$.
A least squares regression gives the prefactor $A\approx 0.17$. The
qualitative features of the numerical simulations are thus in good
agreement with the theory.  Quantitatively, however there is a
discrepency: the theory predicts that we should have $A/z_0=1$, while
our numerical simulation gives $A/z_0\approx 0.6$.  We believe that
the discrepancy arises because the convergence of our numerical scheme
requires we keep a small $\epsilon \approx 0.1$, while the analysis in
the previous subsection assumes $\epsilon=0$.

\subsubsection{Point Singularity}

\begin{figure}[p]
\centerline{\epsfysize=3in\epsfbox{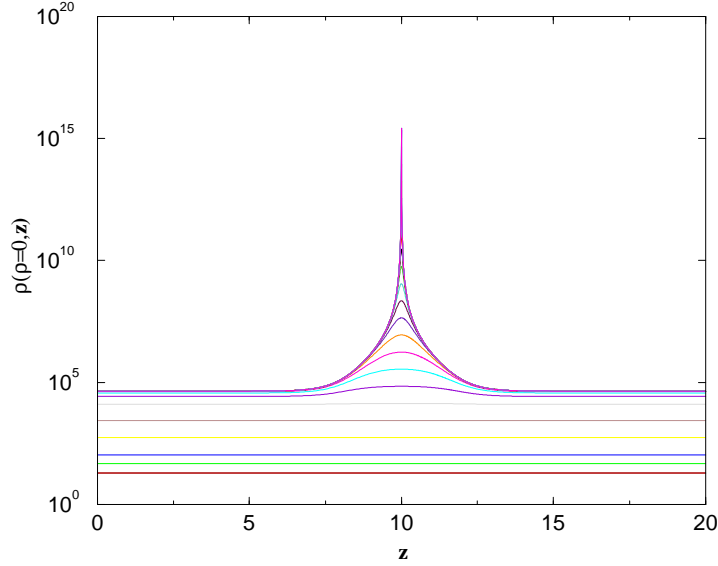}}
\caption[]
{Time evolution of the centerline density $\rho(r=0,z)$, for a point
singularity.  The singularity was initiated by placing a perturbation
(symmetric about $z=10$) on a uniformly collapsing cylindrical
solution.}
\label{prop2}
\end{figure}

Stationary singularities---in which the blow-up happens at a spatial
point---occur in the numerics for a wide variety of initial
conditions. We believe that this represents the generic evolution of a
modulated cylinder. Indeed, it is the end state of the travelling
solution just discussed, when the propagating wave runs into the
reflection--symmetric boundary condition.  An instance of this
solution is depicted in figure \ref{prop2}.

In contrast to the previous case, the corrections to dimensional
scaling are important in
this solution.  For our analysis, we take the point of blow-up to be
at $z=0$. Satisfying the equation requires that the $z$ length scale
incorporate corrections to the scaling:
\[
\tau = \tau_o \phi \left(\frac{z}{\tau_o^{\beta} h(\tau)} \right)
\]
Using $\xi=z/(\tau_o^{\gamma})$
the phase equation becomes 
\[
f^{-2} \left(-\phi - \xi \phi' (\gamma+ \tau_o \frac{h'}{h})\right) = 
\tau_o^{1-2 \gamma} h^{-2} \left(\phi''-\frac{\phi'^2}{\phi}\right)
\]
Demanding that the two sides scale the same way in time (and assuming
$h$ has a power law form, so that $ \tau_{ o} h'/h = $constant), we
have
\[
\frac{1}{f^2}=\frac{\tau_o^{1-2 \gamma}}{h^2}
\]
which gives $\gamma=1/2$ and $f=h$. Thus $
L_z \sim \tau_o^{1/2}  \tau_{ o}^{ -3/16}$,
which differs from
the radial length scale $L_r \sim \tau_o^{1/2} \tau_o^{ 3/16}$. The
result is
\begin{equation}
\frac{L_z}{L_r} = \tau_o^{-3/8}
\end{equation}
This shows that the generic density singularity that forms during the
breakdown of cylindrical collapse, is {\sl not} spherical collapse,
but something milder.  Locally, since the axial scale $L_z$ is much
larger than the radial scale $L_r$ the structure still looks like a
cylinder.  Numerical evidence for this conclusion is evident in Fig
13, which shows that the singularity develops a length scale in the
axial direction which is much larger than the radial scale
$1/\sqrt{\rho}$.  For example, in Fig 13 when $\rho=10^{10}$ (so
$L_r=10^{-5}$) then $L_z \sim 10^{-2}$.  Our numerical algorithms have
unfortunately not allowed us to find the asymptotic $L_z/L_r$
numerically; the problem is that the separation of scales is so great
between the radial and axial scales that one needs many more mesh
points than we can afford to resolve the asymptotic regime.

\section{ Breakup into Spherical Aggregates}

The question of relevance to the experiments is what happens next:
Once blowup occurs at a spatial point the cylinder has a ``free end'',
which changes the nature of the collapse. We can no longer use the
strategy of the previous section---perturbation about a collapsing
cylinder---because the radial structure is no longer closely
approximated by the cylindrical solution.  The pinch-off drives the
dynamics; specifically, the sharp end of the cylinder forms a
travelling wave. Heuristically, note that an edge of bacteria produces
a higher concentration of attractant where the density is higher.
Thus the ``tail'' of bacteria moves toward higher attractant density,
and a travelling wave can form.  Quantitatively, recall that for
variation in one spatial dimension the equations reduce to the Burgers'
equation, which has travelling wave solutions\cite {whitham}.  The
contraction of a cylinder end has been observed for the bacteria
\cite{budper}, and the travelling waves have been discussed in other
contexts\cite {bre98}. Here we discuss the instability of the
recoiling end and  the final
spacing of the spheres.

\begin{figure}
\centerline{\epsfysize=1.2in\epsfbox{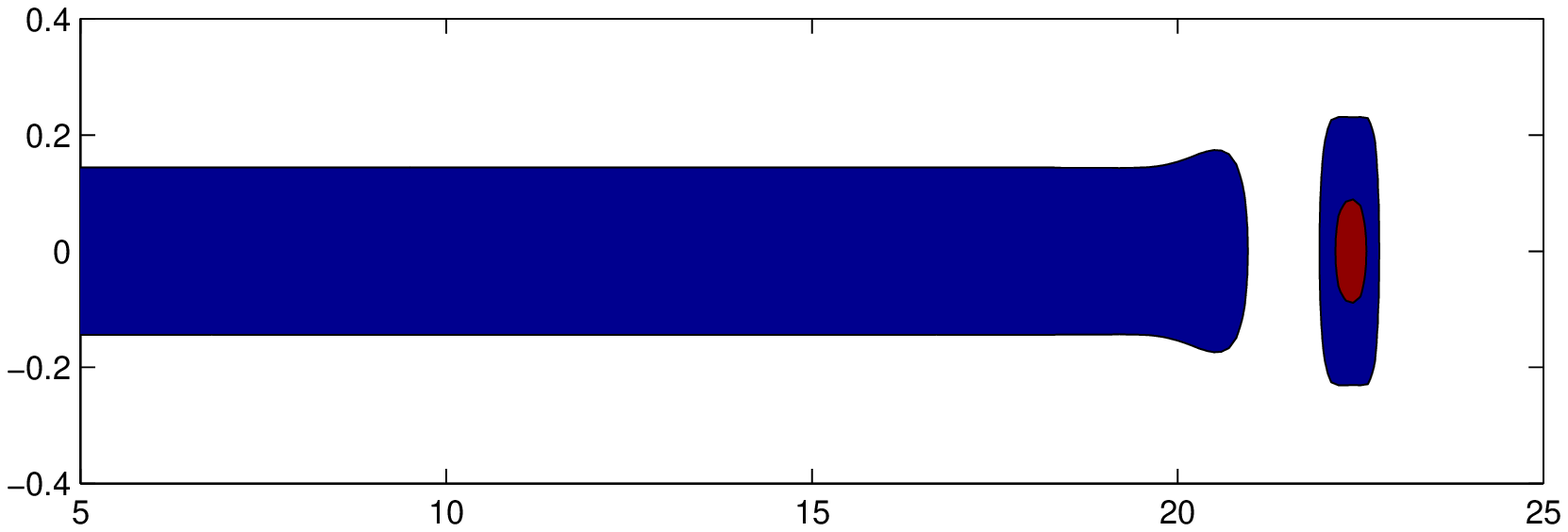}}
\centerline{\epsfysize=1.2in\epsfbox{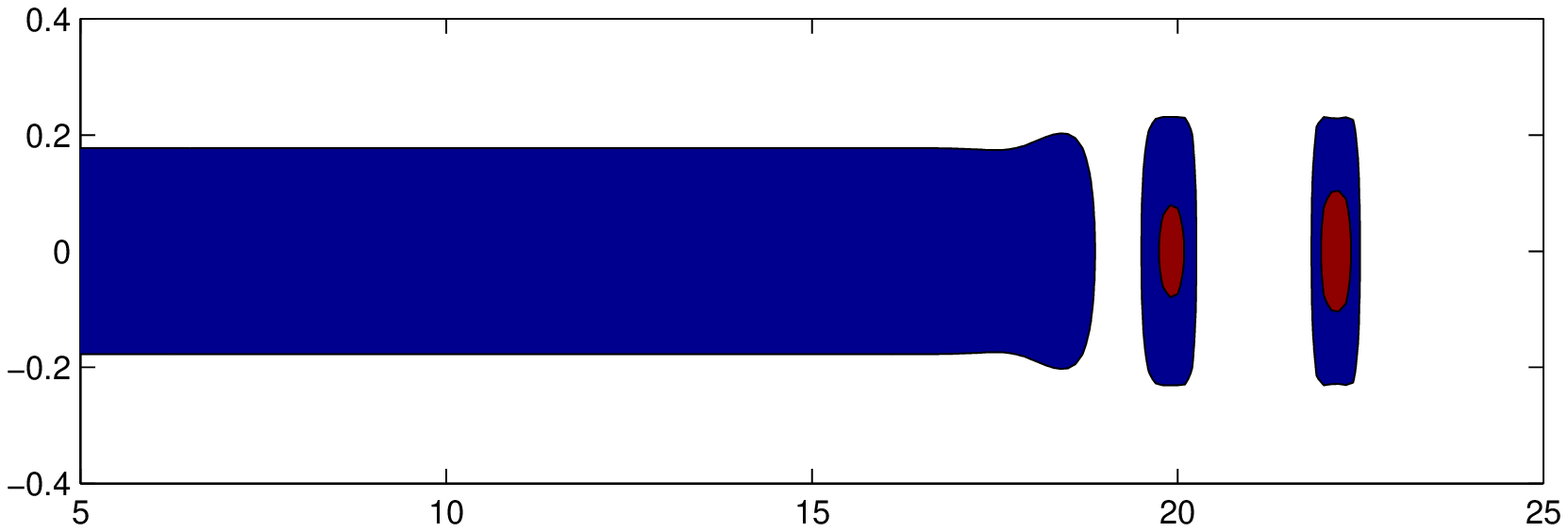}}
\centerline{\epsfysize=1.2in\epsfbox{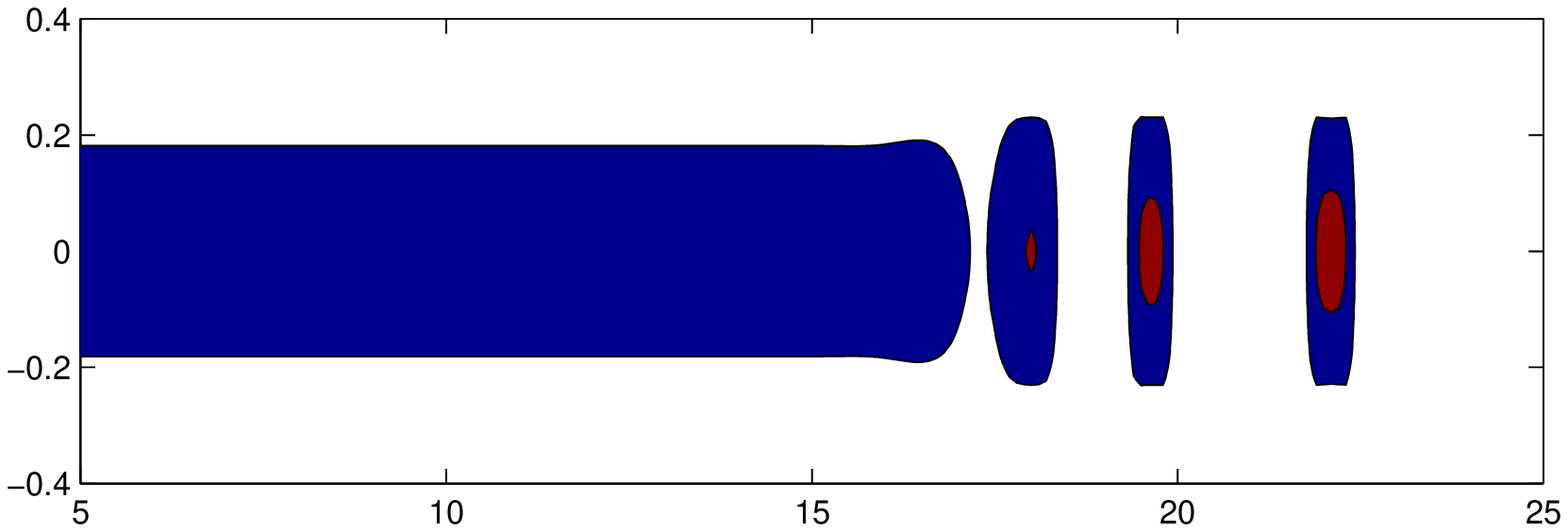}}
\centerline{\epsfysize=1.2in\epsfbox{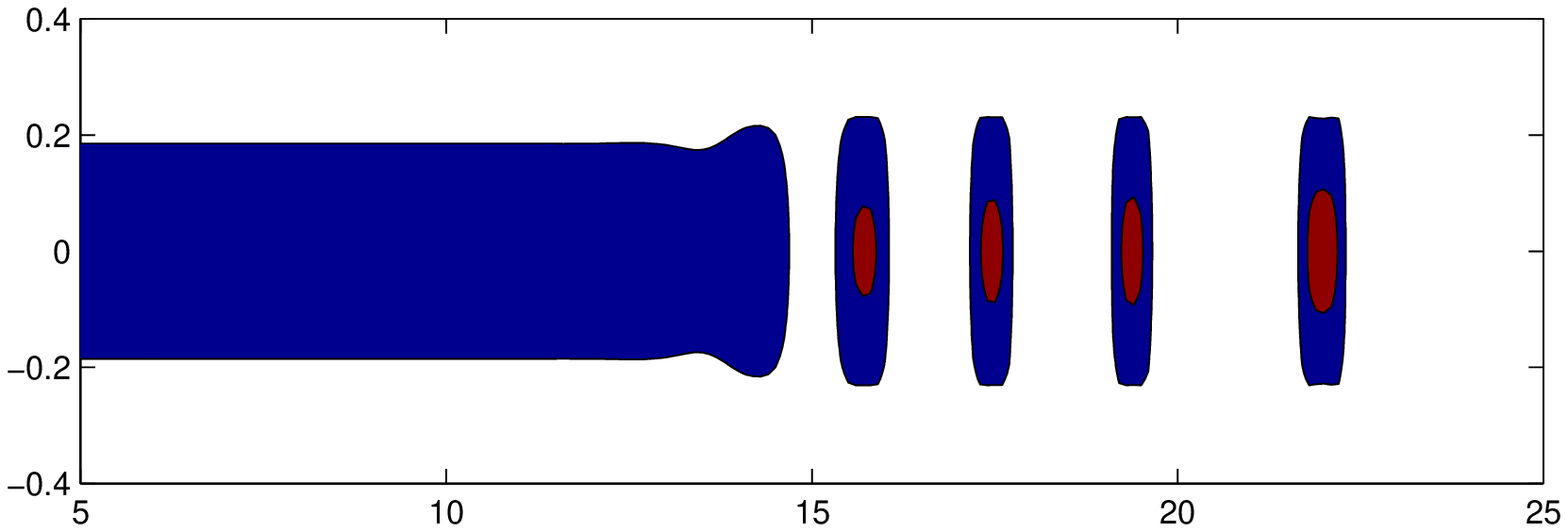}}
\centerline{\epsfysize=3in\epsfbox{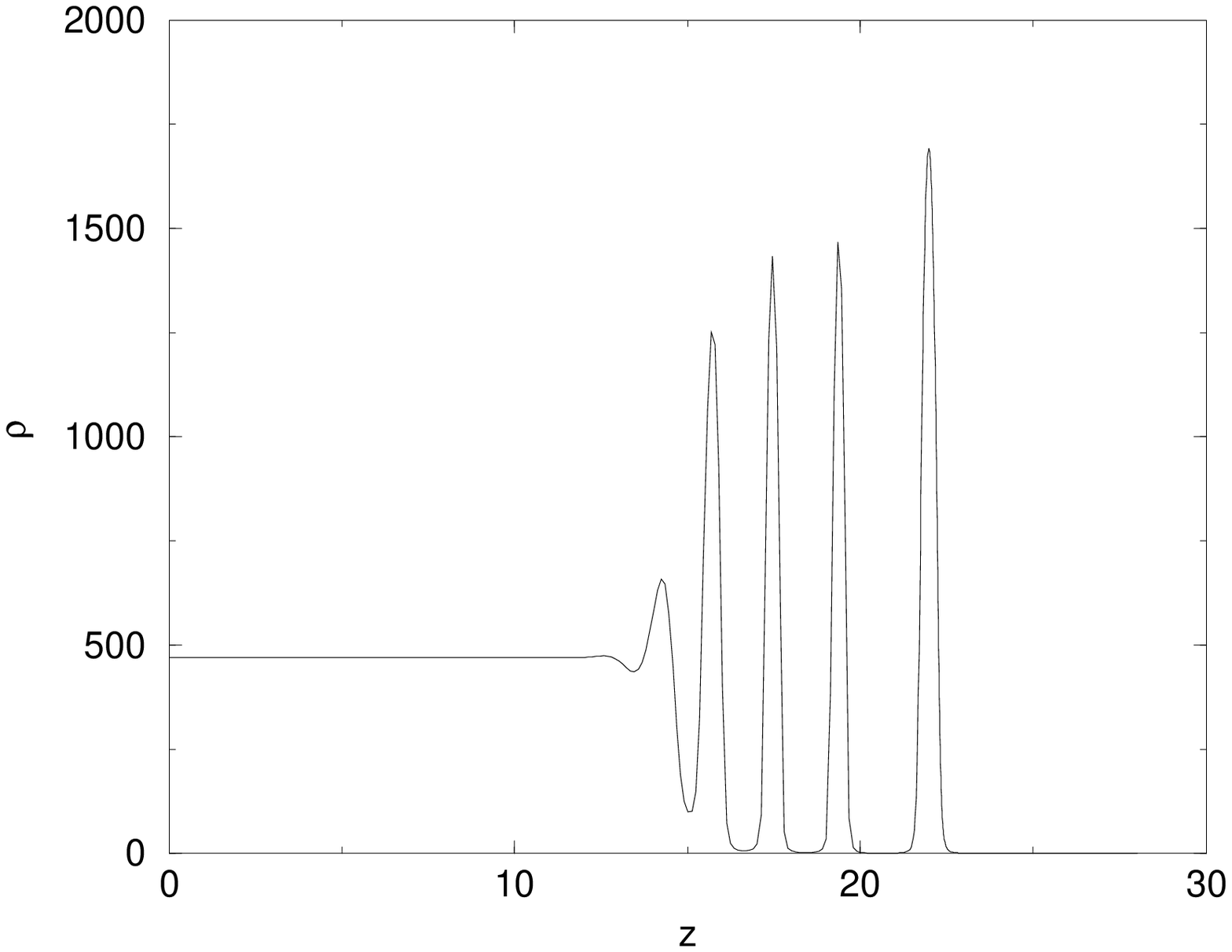}}
\caption[]
{Numerical simulation of a collapsing cylinder with a free end,
showing contour plots of the density.  As the cylinder recoils,
aggregates are left behind.  The spacing between the aggregates is
determined by the density of the cylinder as it collapse.  The final
frame in this figure shows the density profile along the centerline of
the cylinder for the last contour plot (with 4 aggregates).  Note the
maximum density is near the cutoff density $\rho_*=500$ discussed in
the text.}
\label{cylend}
\end{figure}

Figure \ref{cylend} shows a set of numerical simulations of a
retracting cylinder.  The simulation shows that the ``end'' of the
cylinder collapses into a spherical aggregate, and simultaneously, in
front of the aggregate ``waves'' travel into the bulk of the cylinder.
In the simulations, the ends of the cylinder actually try to collapse
to an aggregate of infinite density. In order to continue beyond this
singularity and simulate the formation of an array of aggregates we
introduced a cutoff which emulates the biochemistry in the actual
experiment: when the bacterial density becomes too high, the
bacteria consume all the food sources (succinate and oxygen) in their
local environment and cease producing the attractant aspartate.  We
modelled this by changing the equation for $c$ to
\[
\frac{\partial c}{\partial t} = D_c \nabla^2 c + \alpha \rho \: 
e^{-\rho/\rho^* }
\]
Here we have introduced $\rho^*$, the cutoff
bacterial density.  In \cite{bre98} this cutoff density was estimated
(assuming the cutoff was caused by oxygen depletion) and shown to vary
exponentially with the overhead oxygen concentration in the cell:
$\rho^* \sim e^{C_{Ox}}.$ For the simulation shown in Figure \ref{cylend} this
cutoff is $\rho^*=500$.  Note that the density of
the (undisturbed) cylinder in front of the retracting rim slowly
approaches the cutoff density $\rho^*$; we have found in simulations
that the
undisturbed cylinder always collapses to a density close to the cutoff
value.

The formation of the density wave occurs because the retracting end
perturbs the cylinder in front of it. We can find
 the time-evolution of perturbations to the cylinder, requiring that they
 decay away from the free end. The most unstable mode can be found using
 methods of stationary phase(cf. \cite{lif}).

If the linear growth rate is $\omega(q)$ the point of stationary phase
$q_*$ satisfies \cite{lif}
\begin{eqnarray*}
\mbox{ Im}\: \frac{d \omega}{dq}|_{ q_{*}}  & = & 0 	\\
\mbox{ Re}\:  \frac{d \omega}{dq}|_{ q_{*}} & = & 
\frac{\mbox{ Re}(\omega)}{ \mbox{Im}(q)}
\end{eqnarray*}

For the discussion here, we perform the calculation using the free
space dispersion relation.  A perturbation to constant density has the
form
\begin{eqnarray*}
\rho &=& \rho_o + \delta e^{\omega t - q z}	\\
c &=& \rho_o t+ \chi e^{\omega t - q z}
\end{eqnarray*}
where we have taken $ \epsilon = 1 $ for simplicity in this
calculation.  Plugging into the equations and linearizing gives the
dispersion relation:
\[
\omega =- i \sqrt{\rho_0 } q + q^{ 2}
\] 
The most unstable mode is, in dimensionless units,
\begin{eqnarray*}
v_{*} & = & \sqrt{\rho_0 } \\ 
q_{*} & = &\frac{ \sqrt{\rho_0 }}{ 2} (1
\mp i)\\ 
\omega_{*} & = &\frac{\rho_0 }{2}.
\end{eqnarray*}
These formulae demonstrate that the wavelength of the modulations are
governed by the undisturbed density in front of the rim.  Since this
density is asymptotically determined by the {\it cutoff} $\rho_*$, it
follows that the wavelength of the ripples is determined by the
cutoff. The characteristic distance between the
aggregates is, therefore, determined by the cutoff.  This conclusion
can be experimentally tested.

The predictions for the wavelength and velocity of the front compare
well with numerical simulations. On decreasing $\rho_*$ from $500$ to
$140$ the wavelength of the ripples increases from $\approx 1.4$ to $2.5$, in
qualitative agreement with the formulas.

We remark that the basic scenario outlined in this section was
discovered by Elena Budrene, in unpublished experiments: After
observing the travelling band to collapse as a cylinder (Fig. 1)
Budrene observes fast ``waves'' propagating around the cylinder.
Then, a fragmentation front moves across the collapsing cylinder,
leaving spherical aggregates behind.  The present theory predicts a
scenario that is at least qualitatively very similar: the fast
``waves'' correspond to the excitations of the cylinder from the phase
equation (i.e. the travelling steps, described in section 3).  The
fragmentation front occurs due to the mechanism outlined in this
section.  Unfortunately, it is not currently possible to make a
quantitative comparison of the experiments to the present theory,
though such a comparison would prove most interesting.
  
\section{Connection to Experiments}

This paper has shown how the patterns formed by {\it E. coli} are
connected to the geometry of singularity formation in the hydrodynamic
description of the bacteria. We have constructed the solution for
critical (two-dimensional) collapse, and developed a theory for
modulations to the cylinder. The phase equation provides a useful
simplified description of a perturbed cylinder. Ultimately, the
spacing of spherical aggregates is determined by the instability of a
pinched cylinder of bacteria.

Here we compare our work to published experiments and suggest tests of the
theory.  Not all the coefficients in the original equations have been
precisely measured \cite{bre98} for the experimental
regime of interest.  In particular, neither the attractant
production rate $ \alpha $ nor the chemotactic coefficient $k $ have
been measured for bacteria in the same chemical environment as that of
the collapse experiments.  Thus, at this stage we can make only order
of magnitude numerical comparison with experiments.  Here we use the
values of the coefficients for bacteria in a liquid medium
\cite{bre98}: bacterial diffusion coefficient $D_{ b} = 7 \mbox{x} \
10^{ -6} \mbox{cm}^2$/sec; attractant diffusion coefficient $D_{ b} =
10^{ -5} \mbox{cm}^2$/sec\cite{algebra}; chemotactic coefficient $k = 10^{ -16}
\mbox{cm}^5$/sec; and attractant production rate $ \alpha = 10^{ 3}
$/second/bacteria

An important prediction of this theory is the critical mass of
bacteria for cylindrical collapse: assuming the above parameters, the
formation of a collapsing cylinder requires a minimum number of
bacteria per unit length $M = 4D_{ b} D_{ c}/(k \alpha) =3 \mbox{x}
10^{ 3} $/cm.  The existence of a critical number of bacteria for
cylindrical collapse has been inferred from experiments\cite{bre98},
but this number has never been directly measured.  We emphasize that
(with precise experimental measurements for the parameters $D_b$, $k$
and $\alpha$) the theory rigorously and precisely predicts this
critical mass, allowing a direct test of the theory.

In this paper we have extensively discussed how to construct the
correct description of two dimensional collapse.  In the experiments,
the subtle corrections to the dimensional scaling are probably not
directly observable.  However, the basic scaling relations expected from
the similarity solution---for example, that the maximum density is
related to the length scale of density variations by $ \rho_{ m} \sim
L^{ -2}$--- could be measured in experiments, both for cylindrical and
spherical collapse.  So far, no quantitative and controlled
measurements of the bacterial density have been performed.

In the section on cylindrical collapse, we discussed both the
``transient'' initial regime, and the final asymptotic regime which
appears after true scale separation between the inner and outer
solution is achieved.  In practice, we expect that the experiments
probe only the transient regime.  The hard upper limit on bacterial
density is when the bacteria are closely packed; this corresponds to a
density of $2\mbox{x} 10^{ 9}/\mbox{cm}^3$.  A typical initial density
of bacteria in the liquid experiments\cite{bre98} is
$10^{6}/\mbox{cm}^3$.  The density increases by at most three orders
of magnitude during the collapse, in contrast with the ten orders of
magnitude it takes to reach the asymptotic regime in our simulations.

To our knowledge, modulations to a collapsing cylinder have never been
quantitatively measured in experiments (although as mentioned
above, Budrene has made qualitative observations of this effect).  
The travelling and point
singularities that we predict for a modulated cylinder may be
observable.  In particular, we predict that the radial and axial
length scales should be different for the point singularity.  Because
the difference in these length scales arises directly from the
corrections in two-dimensional collapse, a measurement of these length
scales would test the validity of our derivation of the
two-dimensional solution.

The modulated cylinder ultimately pinches off a point.  We have argued
that the spacing of spherical aggregates is determined by the
instability of a cylinder with an end.  In practice, when does the
modulated cylinder pinch off (forming an end)?  To answer this
question, we must know when our theory breaks down.  Collapse to
infinite density cannot happen for bacteria, because they have finite
size.  It was argued in
\cite{bre98} that even before the hard packing density of bacteria is
reached, oxygen depletion will stop bacterial collapse.  Regardless of
the specific mechanism for stopping the collapse, at some time the
highest density part of the cylinder---the 
point singularity---will stop collapsing.  This is the time
of pinch off:  the point singularity 
evolves much more slowly than the neighboring, less dense regions of
the cylinder.

This argument about the point of pinch off gives a testable prediction
of our model, because the spacing of aggregates depends on the maximum
density of the cylinder.  In dimensional units, the most unstable
wavelength (and aggregate spacing) is
\[
 \lambda = 4 \pi \sqrt{\frac{D_b D_c}{ \alpha k}} \rho_{ m}^{ -1/2}
 =300 \mbox{cm}^{-1/2} \rho_{ m}^{ -1/2}
\]
where we have used values of the coefficients from above.  Thus,
varying the maximum attainable density of the bacteria should cause
the aggregate spacing to change according to this scaling law.  In
\cite{bre98} a formula for how the maximum density in a collapsed
aggregate depends on the oxygen concentration $C_{Ox}$ was derived,
and shown to be $\rho_m \sim e^{C_{Ox}}$. This implies that the
wavelength of the pattern should decrease exponentially with the
oxygen concentration; systematic experiments could test this
prediction.

The one solid prediction that can be compared with present experiments is that
there is a {\sl lower bound} on the aggregate spacing, that follows
from the hard-packing
density of bacteria.  Using the characteristic
size $10\mu$m of {\it E. coli} this is approximately
$10^9 /\mbox{cm}^3 $.  Thus the measured aggregate spacing should always be above
the lower bound
\[
 \lambda _{ \mbox{min}} = 0.1  \mbox{ mm},
\]
where in this estimate we used the assumed values for the constants $D_b,k$ and
$\alpha$.  This lower bound agrees with experiments, in that the spacings
are typically measured in millimeters.

{\bf Acknowledgements:} We are grateful to Elena Budrene both for
teaching us about her experiments that inspired this work and for many
useful discussions.  We  acknowledge early discussions with Leonid
Levitov, Leo Kadanoff and Shankar Venkataramani, and thank Daniel Fisher for
helpful comments.  This research was supported by the National Science
Foundation Division of Mathematical Sciences, and the A. P. Sloan
Foundation. MDB acknowledges support from the Program in Mathematics
and Molecular Biology at the Florida State University, with funding
from the Burroughs Wellcome Fund Interfaces Program.

\appendix

\section{Remarks on Numerical Methods}
\label{numerics}

The partial differential equations described in this paper were solved
using second order in space, finite difference methods, supplemented
with adaptive mesh refinement.  The time discretization used a
$\theta$ weighted Crank-Nicolson-type scheme (i.e. in the equation
$\dot{f} = {\bf L} f$ the right hand side is evaluated at time
$(n+\theta)\Delta t$, where $\Delta t$ is the timestep).  Typically, in
the simulations with one spatial dimension, $\theta=0.6$.  For the
simulations in two spatial dimensions, we used an ADI operator
splitting method, which requires using $\theta=1$.  Because these
methods are implicit, at each timestep a matrix inversion was
necessary.  This is the most expensive part of the numerical method.

The most subtle aspect of the numerical simulations reported in this
paper is the mesh refinement.  Without good mesh refinement it is
impossible to get close enough to the singularity to resolve the
(logarithmic) corrects uncovered for the cylindrical collapse; without
good mesh refinement in the two-dimensional simulations it would be
impossible to acquire enough decades of data to test the phase
equation-theory presented in section 4. The mesh refinement was
performed as follows:

{\sl One Spatial Dimension:} The philosophy of mesh refinement
employed in this paper (first explained to us by Jens Eggers) is to
frequently implement gradual changes in the mesh, as opposed to
infrequently implementing large changes.  Mesh refinement is
implemented every time the maximum density increases by one percent.
During the refinement, the characteristic scale over which the
solution varies is determined, and a mesh is constructed to well
refine this scale; typically, this involves making sure there are
about one hundred mesh points across the region that the solution
varies significantly.  The solution at the new mesh is constructed by
cubic spline interpolation of the old mesh.  Because the mesh is
refined frequently, the changes to the solution occurring during
refinement are small, and there are no convergence difficulties after
refinement.  The algorithm allows finding solutions over essentially
arbitrary changes in the bacterial density with as little as few
hundred mesh points.  The simulations reported in the paper typically
use a few thousand mesh points, in order to accurately compute the
slow approach to the {\rm log log} scaling regime.

{\sl Two Spatial Dimensions:} In two dimensions, the equations are
solved using standard operator splitting techniques.  The mesh is
rectangular, described by two functions, the $x$ coordinate $x_i$ and
the $y$ coordinate $y_j$.  Because the one dimensional algorithm
described above can achieve arbitrary resolution with a few hundred
mesh points, it is possible to well resolve density changes in both
spatial directions using of order $10^4$ mesh points.  The algorithm
for these simulations works analogously to that for the one spatial
dimension case described above: Every fifty time steps, the $x$ grid
(or $y$ grid) is remeshed, in accordance with the criterea outlined
above.  Typically we stagger the remeshing between the two directions
by twenty five timesteps.

Both the one and two dimensional codes were tested extensively by
checking the solutions against known analytical results.  All of the
results that are presented in this paper follow the philosophy that
numerical results are only believable if they can be replicated by
asymptotic solutions of the equations; in turn, asymptotic results are
only useful if they show up in numerics.  For the two-dimensional
code, one might worry that the operator splitting coupled with the
remeshing induces artifical biases in the numerics; besides checking
our numerical solutions against solutions to the phase equation, we
have also tested the two dimensional code by checking that it can
reproduce the scalings and the similarity solution for spherically
symmetric collapse, where the solution is known very well.

\section{Phase Equation for a Bacterial Cylinder }
\label{phasedet}

In this section we fill in the details of the calculation of the phase 
equation.
Evaluating the coefficients in equation (\ref{phaseco}) we find\cite{coeffcor}
\begin{eqnarray*}
F(R_o,C_o) &=& R_o + \frac{\eta R_o'}{2}			\\
G(R_o,C_o) &=& R_o + \frac{\eta R_o'}{2}	 -\frac{\eta R_o C_o'}{ 2}	\\
H(R_o,C_o) &=&2 R_o + \frac{1}{4} (7 \eta R_o' + \eta^2 R_o'' - 5 \eta R_o
C_o' - \eta^2 (R_o C_o')' )  
\end{eqnarray*}

To evaluate the solvability condition, we need to find the zero mode
of the adjoint to the linearized operator. In this case, the linear
operator is the matrix
\[
\left(	\begin{array}{cc}
\nabla^2 - \nabla \cdot (\nabla C_o \cdot)& - \nabla \cdot( R_o \cdot) 	\\
1	&		\nabla^2  
\end{array}		\right)
\]
All of the terms in $\Lambda$ are self-adjoint except those of the
form $\nabla \cdot (\nabla C_o \cdot)$. Under the definition (for a
cylinder of bacteria) of the
inner product $\langle f,g \rangle = \int r dr \int dz f^* g$ we can
determine the adjoint
\[
[\nabla \cdot( \nabla C_o \cdot)]^{\dagger} = - \partial_{r} C_o
\partial_{r}  
\]
which gives the adjoint linear operator 
\[
\Lambda ^{\dagger} = 
\left(	\begin{array}{cc}
	\nabla^2 +  \nabla_{r} C_o \nabla_{r} &  1 \\
	-\nabla \cdot( R_o \nabla \cdot)	&	\nabla^2 
\end{array}		\right).
\]
This linear operator possesses a simple zero mode: $\Lambda ^{\dagger}
(1,0)=0$. The coefficients of phase equation thus become
\begin{eqnarray*}
c_1 &=& \langle 1, F(R_o,C_o) \rangle	\\
c_2 &=& \langle 1, G(R_o,C_o) \rangle 	\\
c_3 &=& \langle 1, H(R_o,C_o) \rangle 
\end{eqnarray*}

A subtlety comes when we evaluate the inner products: we must
integrate (in similarity variables) to the upper limit of validity of
the similarity solution. For a cylinder, this upper limit is $\eta_*$, 
the radius at which the solution matches onto the outer solution. From 
the asymptotics discussed earlier, we use that $\eta_*\sim A^{-1/2} =
f(s)$. Evaluating  the inner products, and taking 
the limit $\tau \rightarrow  0$, we arrive at the result
\begin{eqnarray*}
c_1 &=& \frac{-4}{f^{2}}	\\
c_2 &=& 4					\\
c_3 &=& -4
\end{eqnarray*}
Which gives as the phase equation
\[
\frac{\tau_t+1}{f^2} =\tau_{zz}
-\frac{\tau_z^2}{\tau} 
\]


\begin{thebibliography}{10}
\bibitem{simulation1}
The simulation shown here used a box size $W = 40 $ with a total mass of $M =
  12 $ and $ \epsilon = 10^{ -5} $. For boundary conditions, we demanded
  reflection symmetry at the origin and zero flux at the outer boundary.
  Different initial conditions give similar final results; here we used a
  Gaussian profile.
\bibitem{transient}
In principle, the transient regime can be eliminated or shortened by preparing
  the system with mass very close to the critical mass $M = 4 $. With less
  excess mass to expel, the system approaches the asymptotic state more
  quickly. However, even in simulations with $M \approx 4 $ the transient
  regime still persists for several decades. Since the density can increase by
  at most three or for orders of magnitude in experiments, the transient regime
  dominates the observed behavior.
\bibitem{simulation2}
This simulation used $W=10$, mass $M=12$, and $\epsilon=10^{-5}$. The
  characteristic length scale of the density distribution was varied, between 5
  and $10^{10}$, as shown in the Figure.
\bibitem{boxsize}
Here we note an interesting feature we observe in simulations: because the
  features of the transient regime are determined by matching to the boundary,
  the crossover time $\tau_c$ when the system reaches the true asymtotic regime
  varies with the system size. For a larger box size, the outer ``bump'' is
  farther away (in similarity variables) at a given time, and therefore the
  asymptotic regime is reached sooner.
\bibitem{boundarycondition}
Note that the presence of the bump reflects the boundary conditions applied in
  the simulation: we used no flux boundary conditions. As an example of a
  different boundary condition, requiring constant density at the boundary
  means the bump will not be present.
\bibitem{innerproduct}
The inner product here is $\langle f,g \rangle = \int r \: dr \int dz \: f^*
  g$.
\bibitem{simulation3}
For the simulations presented in this section, the cylinder was represented on
  a mesh in $(r,z)$, with boundary conditions that $ \rho(W)=0$ and $c(W)=0$,
  with $W$ a fixed constant ($W=5$ in the simulations below). Reflection
  symmetry was enforced at both $z=0$ and $z=L=20$, so that the solution is
  periodic in $z$.
\bibitem{algebra}
Note that the estimate for the chemotactic coefficient given in \cite{bre98}
  has an algebraic error.
\bibitem{coeffcor}
The values of coefficients here were computed with $f(\tau)=1$ for simplicity.
  When $f$ varies in time, there will be slight differences in th e numerical
  values of the coefficients; however these coefficients remain consta nt in
  time and the qualitative behavior of the phase equation is unchanged.
\end{thebibliography}

\end{document}